\documentclass[twocolumn]{aastex62}
\usepackage[utf8]{inputenc}
\usepackage{amsmath}
\hypersetup{linkcolor=cyan,citecolor=cyan,filecolor=cyan,urlcolor=cyan}
\usepackage{color}
\definecolor{mygreen}{HTML}{139703}

\newcommand{\ms}{\ensuremath{\mathrm{m\,s}^{-1}}}
\definecolor{wynn_red}{HTML}{E74C3C}

\defcitealias{jphc11}{JPH11}

\begin{document}
\title{Measuring the Orbital Parameters of Radial Velocity Systems in Mean Motion Resonance---a Case Study of HD 200964}
\author{M. M. Rosenthal}
\affiliation{Department of Astronomy and Astrophysics, University of California, Santa Cruz, CA 95064, USA}
\author{W. Jacobson-Galan}
\affiliation{Department of Astronomy and Astrophysics, University of California, Santa Cruz, CA 95064, USA}
\author{B. Nelson}
\affiliation{Redwood Center for Theoretical Neuroscience, University of California,
Berkeley, Berkeley, CA 94720, U.S.A.}
\author{R. A. Murray-Clay}
\affiliation{Department of Astronomy and Astrophysics, University of California, Santa Cruz, CA 95064, USA}
\author{J. A. Burt}
\affiliation{Department of Physics and Kavli Institute for Astrophysics and Space Science, Massachusetts Institute of Technology, Cambridge, MA 02139, USA}
\author{B. Holden}
\affiliation{UCO/Lick Observatory, Department of Astronomy and Astrophysics, University of California at Santa Cruz, Santa Cruz, CA 95064, USA}
\author{E. Chang}
\affiliation{Department of Astronomy and Astrophysics, University of California, Santa Cruz, CA 95064, USA}
\author{N. Kaaz}
\affiliation{Department of Astronomy and Astrophysics, University of California, Santa Cruz, CA 95064, USA}
\author{J. Yant}
\affiliation{Department of Astronomy and Astrophysics, University of California, Santa Cruz, CA 95064, USA}
\author{R. P. Butler}
\affiliation{Department of Terrestrial Magnetism, Carnegie Institution for Science, Washington, DC 20015, USA}
\author{S. S. Vogt}
\affiliation{Department of Astronomy and Astrophysics, University of California, Santa Cruz, CA 95064, USA}
\affiliation{UCO/Lick Observatory, Department of Astronomy and Astrophysics, University of California at Santa Cruz, Santa Cruz, CA 95064, USA}
\begin{abstract}
The presence of mean motion resonances (MMRs) complicates analysis and fitting of planetary systems observed through the radial velocity (RV) technique. MMR can allow planets to remain stable in regions of phase space where strong planet-planet interactions would otherwise destabilize the system. These stable orbits can occupy small phase space volumes, allowing MMRs to strongly constrain system parameters, but making searches for stable orbital parameters challenging. Furthermore, libration of the resonant angle and dynamical interaction between the planets introduces another, long period variation into the observed RV signal, complicating analysis of the periods of the planets in the system. We discuss this phenomenon using the example of HD 200964. By searching through parameter space and numerically integrating each proposed set of planetary parameters to test for long term stability, we find stable solutions in the 7:5 and 3:2 MMRs in addition to the originally identified 4:3 MMR.  The 7:5 configuration provides the best match to the data, while the 3:2 configuration provides the most easily understood formation scenario. In reanalysis of the originally published shorter-baseline data, we find fits in both the 4:3 and 3:2 resonances, but not the 7:5. Because the time baseline of the data is less than the resonant libration period, the current best fit to the data may not reflect the actual resonant configuration. In the absence of a full sample of the longer libration period, we find that it is of paramount importance to incorporate long term stability when fitting for the system’s orbital configuration.
\end{abstract}

\section{Introduction}
A $p$:$q$ mean-motion resonance (MMR) occurs when the ratio of the periods of two interacting planets is close to $p/q$. This commensurability allows planetary conjunctions to occur at consistent locations in the planets' orbits, leading to periodic transfers of energy and angular momentum between the two bodies.  Many examples of bodies in mean motion resonance are known in the solar system (for a review, see e.g. \citealt{p_1986}) and in exoplanetary systems (e.g.  \citealt{lff_2011}, \citealt{ior_2017}).  In this paper, we restrict our focus to systems of giant planets in MMR. Mean motion resonance between Jupiter and Saturn has been suggested as a possible phenomenon early in the solar system's history (e.g. \citealt{mtc_2007}, \citealt{wmr_2011}). Several sets of giant planets in resonance have been identified directly (e.g., GJ 876, \citealt{lp_2002}; HD 5319, \citealt{gfp_2015}; HD 33844, \citealt{wjb_2016}; HD 47366, \citealt{mwh_2018}; HD 202696, \citealt{tsh_2019}; and TOI-216, \citealt{knh_2019}) and resonance has been inferred due to stability constraints in the directly imaged system of giants HR 8799 \citep[e.g.,][]{fmc_2010,wgd_2018}. Understanding the population of giant planets in MMRs is important for constraining the typical migration histories of giant planets, as convergent migration of giant planets in a gas disk is a commonly cited mechanism for formation of gas giants in MMR (e.g. \citealt{lp_2002}). 

Resonances often constitute stable regions in otherwise unstable parts of phase space. Because the interactions between planets in MMR can generate periodic oscillations of the system's line of conjunctions, they can protect planets from close encounters. Thus, MMRs are often invoked to explain observed systems that initially appear to be unstable.

Unfortunately, the presence of MMRs greatly complicates analysis of RV systems. Strong planet-planet interactions cause the planets to deviate from pure Keplerian motion even on the timescale of typical RV observations. This complicates the usual RV fitting process, where planets are often allowed to move on unperturbed Keplerian orbits. Furthermore, the additional frequencies introduced by these dynamical interactions can shift the peaks in a periodogram of the RV signal away from the true orbital periods of the planets. This difficulty in identifying the periods of the planets in turn means that, perhaps counterintuitively, the particular resonance that a system is in is not clear from the outset of fitting. Further exacerbating this issue is the fact that libration of the MMR's resonant angle occurs on timescales that are generally longer than the timescale of the RV observations, meaning that our observations only capture part of the full libration.  This sampling issue, along with error in the observations, means that the best-fit solutions to RV signals may lie far from solutions that actually exhibit long term stability. 

Thus, fitting RV systems in MMR necessitates different methods than those traditionally used to fit radial velocity systems. Firstly, theoretical radial velocities must be generated through full numerical integration of the equations of motion of the system (e.g., \citealt{tpl_2013}, \citealt{wtl_2014}, \citealt{nfw_2014}, \citealt{tkz_2017}, \citealt{mlt_2018}). Furthermore, while initial searches through parameter space can be performed without incorporating long term stability, the ``true" posterior distribution of the planetary orbital parameters should not include points that are unstable on short timescales. In some cases ``rejection sampling", i.e. throwing out all points that do not exhibit stability, can produce posterior distributions conditioned on long-term stability. However, as will be seen in this work, it is often the case that the fraction of stable points is so small that the posterior produced by rejection sampling does not adequately represent the underlying probability distribution.  Thus, in order to find long-term stable posterior distributions it is often necessary to incorporate stability during the search through parameter space, though this is often not explicitly done. Incorporating long term stability makes exploring the parameter space difficult, as while regions close to particular MMRs will exhibit long term stability, intermediate regions will generally have no stable solutions, meaning that each proposed resonance must be investigated separately. 

In this work, we illustrate these difficulties and ways they can be mitigated through the example of the planetary system orbiting the star HD 200964. HD 200964 is an intermediate mass subgiant (see Table \ref{tab:stell_pars} for a summary of the stellar parameters), which was reported by \citet{jphc11} (hereafter \citetalias{jphc11}) to host two massive ($M_p \gtrsim M_J)$ giant planets in a tight orbital configuration ($P_c$:$P_b$ $\sim$800:600 days). \citetalias{jphc11} gave a best-fit, long term ($>10^7$ years) stable solution that was close to a 4:3 MMR. In this work, we include additional observations from both the Keck telescope as well as the Automated Planet Finder (APF), which increase the length of time spanned by the RV data. In addition, we explicitly require stability in our search over parameter space, which greatly aids in finding regions of parameter space that both fit the data well and exhibit long term stability. We find that, in addition to the 4:3 solution identified by by \citetalias{jphc11}, the system can be fit by both a 3:2 MMR and a 7:5 MMR, with the 7:5 providing the best fit to the measured radial velocity. The presence of multiple plausible MMRs highlights the general difficulty in pinning down MMR in observed radial velocity systems. We also note that if the system is truly in a 3:2 MMR, this would mitigate difficulties in forming the system through convergent migration. 

In Section \ref{obs}, we discuss how our observations of HD 200964 were performed. In Section \ref{prev}, we discuss the results of previous analyses of HD 200964. In Section \ref{methods} we discuss the various methods we employed to find best-fit, long-term stable solutions to the observed radial velocity. In Section \ref{res} we analyze the MMRs that stabilize the best-fit solutions we find. In Section \ref{reanaly} we perform our methodology on the \citetalias{jphc11} dataset and compare our results with theirs, and in Section \ref{third} we discuss the possibility of a third planet in the system. Finally, in Section \ref{conc} we summarize our results and give our conclusions.

\begin{deluxetable}{cc} \label{tab:stell_pars}
\setlength{\tabcolsep}{24pt}
\tabletypesize{\footnotesize}
\tablecaption{Stellar parameters for HD 200964, taken from \cite{bfv_2016}}
\tablehead{\colhead{Parameter}& \colhead{Value}}
 \startdata
 $V_{\text{mag}}$ & 6.48 \\
 Distance [pc] & 72.2 \\
 $T_{\rm{eff}}$ & 4982 \\
 $\log g$ & 3.22 \\
 $[\rm{M}/\rm{H}]$ & -0.1 \\
$\log L \left[L_\odot\right]$ & 1.13 \\
$R_* \left[R_\odot\right]$ & 4.92 \\
$M_* \left[ M_\odot \right]$ & 1.45 \\
Age [Gyr] & 3.3 \\
\enddata
\end{deluxetable}

\section{Observations} \label{obs}

The radial velocity measurements of HD 200964 used in this analysis come from three different facilities: the Hamilton spectrometer \citep{Vogt1987} paired with the Shane 3 m or the 0.6 m Coude Auxiliary Telescope, the HIRES spectrometer \citep{vab_1994} on Keck I, and the Levy spectrometer on the Automated Planet Finder (APF) telescope \citep{vrk_2014}. In all cases, the star's Doppler shifts were measured by placing a cell of gaseous iodine in the converging beam of the telescope, imprinting the stellar spectrum with a dense forest of iodine lines from 5000-6200 \AA\ \citep{Butler1996}. These iodine lines were used to generate a wavelength calibration that reflects any changes in temperature or pressure that the spectrometer undergoes, and enables the measurement of each spectrometer's point spread function. Although each spectrometer covers a much broader wavelength range, 3400-9000 \AA\ for the Hamilton and 3700-8000 \AA\ for HIRES and the Levy, only the iodine rich 5000-6200 \AA\ region was used for determining the observation's RV shift. For each stellar spectrum, the iodine region was divided into $\sim$700 individual 2\AA\ chunks. Each chunk produces an independent measure of the wavelength, point spread function, and Doppler shift. The final measured velocity is the weighted mean of the velocities of all the individual chunks. It is important to note that all RVs reported here have been corrected to the solar system barycenter, but are not tied to any absolute RV system. As such, they are “relative” velocities, with a zero point that is usually set simply to the mean of each dataset. 

We make use of two previously published RV datasets, denoted here as the ``Lick" and ``Keck11" datasets, taken from \citet{Johnson2011} which originally announced the detection of these planets in the context of their intermediate-mass subgiant host star survey \citep{Johnson2006, Peek2009, Bowler2010, Johnson2010}. The Lick data have SNR of $\sim$120 in the center of the iodine region ($\lambda = 5500$\AA) corresponding to an internal uncertainty value of 4-5\ms, while the Keck11 have SNR $\sim$180 in the same area which brings the internal uncertainties down to 1.5-2\ms. 
For additional details on this data, see \citet{Johnson2011}. New to this paper are an additional 50 velocities taken with Keck HIRES and 36 velocities taken with the APF, all obtained as part of the long running LCES Doppler survey \citep{Butler2017} and denoted as ``Keck" and ``APF", respectively. For our HIRES observations the median SNR in the iodine region is 159, corresponding to an average internal uncertainty of 1.4\ms. The APF observations have a median SNR of 101 in the iodine region, which produces an average internal uncertainty of 1.5\ms. These internal uncertainties reflect only one term in the overall RV error budget, and result from a combination of systematic errors from things like properly characterizing the point spread function, detector imperfections, optical aberrations, and under sampling the iodine lines, among others. 

\begin{figure*}[htbp]
	\centering
\includegraphics[width=7in]{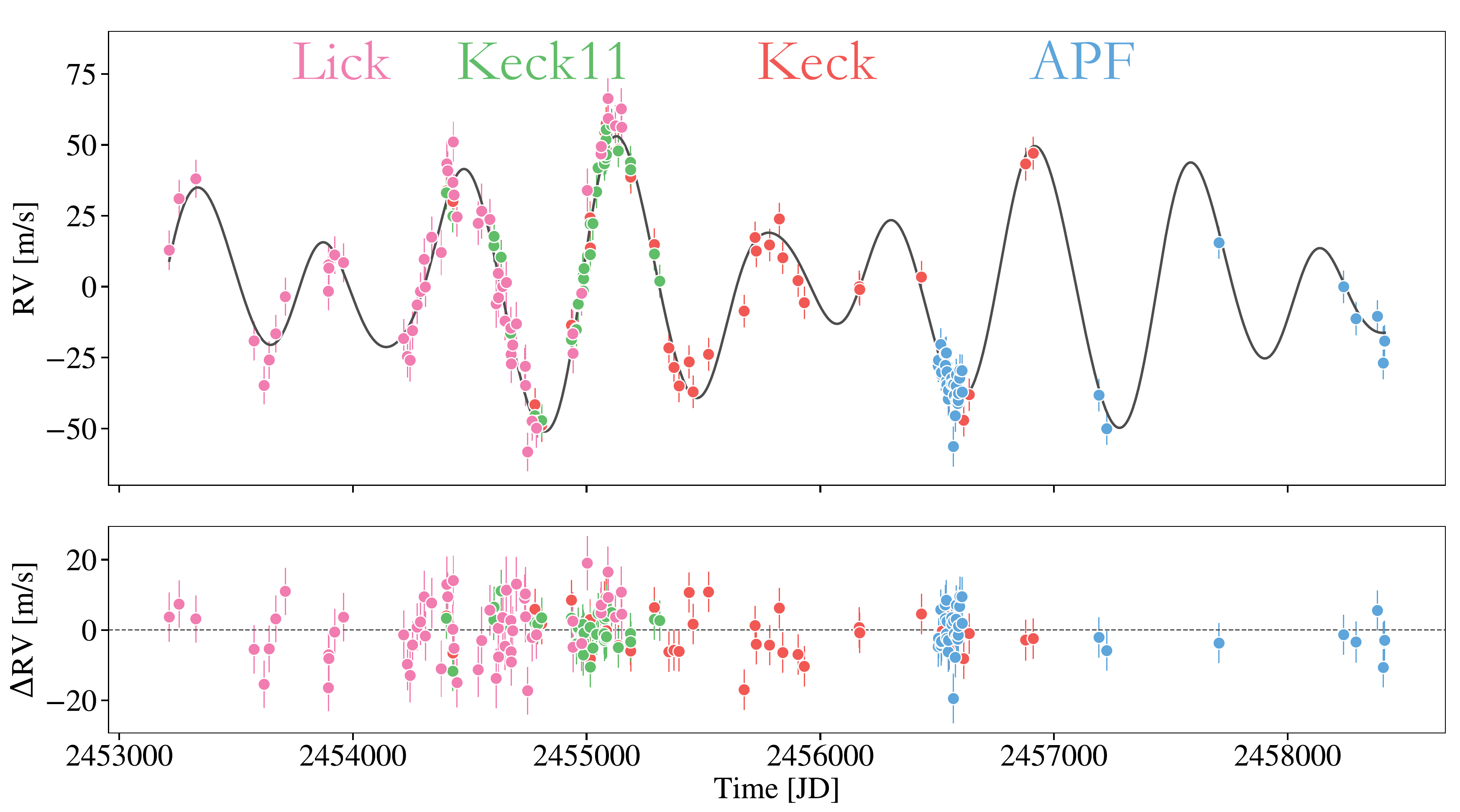}
		\caption{The four data sets for the radial velocity of HD 200964, along with the theoretical radial velocity curve obtained using the parameters given in Table \ref{tab:bestfit_nostab}. The data sets are: Lick (pink points), Keck11 (green points), Keck (red points), and APF (blue points).  Note that, as discussed in Section \ref{fitting}, each data set has a constant offset that we fit separately. Furthermore, the jitter term given in Table \ref{tab:bestfit_nostab} is added in quadrature to the quoted error bars to obtain the error bars shown in the figure. The residuals between the theoretical velocity and the data are shown in the bottom panel.}
\label{fig:rvt_nostab}
\end{figure*}

The new Keck and APF radial velocities are given in Tables \ref{tab:Keck_RV} and \ref{tab:APF_RV} respectively. Additionally, all four data sets, along with our maximum likelihood solution without stability taken into account (see Section \ref{fitting}), are plotted in Figure \ref{fig:rvt_nostab}.

\begin{table} \label{tab:Keck_RV}
\centering
\caption{Keck Radial velocities for HD 200964}    
\begin{tabular}{rrr}
\hline
    Julian Day &        RV [m $
\rm{s}^{-1}$] &      Uncertainty [m $
\rm{s}^{-1}$] \\
\hline
 2454399.75186 &  23.56 & 1.54 \\
 2454427.75745 &  19.83 & 1.13 \\
 2454634.07880 &   0.38 & 1.57 \\
 2454674.91593 & -24.76 & 1.46 \\
 2454778.80262 & -51.78 & 1.54 \\
 2454807.78906 & -59.08 & 1.62 \\
 2454935.13871 & -23.75 & 1.21 \\
 2454956.09772 & -25.09 & 1.39 \\
 2454964.11957 & -16.80 & 1.39 \\
 2454984.06802 &  -4.39 & 1.36 \\
 2454985.09297 &  -5.21 & 1.52 \\
 2455014.96811 &  14.08 & 1.61 \\
 2455015.95302 &   3.45 & 1.56 \\
 2455075.07263 &  40.27 & 1.53 \\
 2455076.06215 &  35.78 & 1.65 \\
 2455077.05115 &  44.44 & 1.51 \\
 2455082.04172 &  35.42 & 1.50 \\
 2455083.04807 &  45.17 & 1.70 \\
 2455084.02263 &  47.35 & 1.48 \\
 2455084.99943 &  38.08 & 1.53 \\
 2455106.90692 &  46.40 & 1.32 \\
 2455135.75335 &  38.24 & 1.40 \\
 2455187.69803 &  30.50 & 1.63 \\
 2455188.69157 &  28.42 & 1.54 \\
 2455290.14918 &   4.66 & 1.47 \\
 2455313.13833 &  -8.29 & 1.12 \\
 2455352.08439 & -31.80 & 1.38 \\
 2455374.11241 & -38.66 & 1.51 \\
 2455395.95755 & -45.18 & 1.38 \\
 2455439.01932 & -36.70 & 1.38 \\
 2455455.73766 & -47.27 & 1.42 \\
 2455521.79477 & -34.05 & 1.39 \\
 2455674.14167 & -18.79 & 1.47 \\
 2455720.97469 &   7.14 & 0.57 \\
 2455726.03586 &   2.36 & 1.28 \\
 2455782.84153 &   4.52 & 1.34 \\
 2455824.92332 &  13.68 & 1.23 \\
 2455839.82582 &   0.00 & 1.27 \\
 2455904.73172 &  -8.06 & 1.43 \\
 2455931.69116 & -15.84 & 1.41 \\
 2456166.74440 & -10.20 & 1.40 \\
 2456168.86382 & -11.20 & 1.21 \\
 2456433.04139 &  -6.80 & 1.14 \\
 2456522.09346 & -38.29 & 1.59 \\
 2456529.87766 & -35.31 & 1.54 \\
 2456551.82347 & -49.97 & 1.18 \\
 2456613.77979 & -57.29 & 1.44 \\
 2456637.69903 & -48.26 & 1.32 \\
 2456878.89942 &  33.07 & 1.56 \\
 2456911.71152 &  36.84 & 1.42 \\
\hline
\end{tabular}
\end{table}

\begin{table} \label{tab:APF_RV}
\centering
\caption{APF Radial velocities for HD 200964}
\begin{tabular}{rrr}
\hline
    Julian Day &        RV [m $
\rm{s}^{-1}$] &      Uncertainty [m $
\rm{s}^{-1}$] \\
\hline
 2456504.82513 &   0.54 & 1.26 \\
 2456505.93027 &   2.60 & 1.29 \\
 2456515.85262 &   8.11 & 1.20 \\
 2456516.89268 &  -2.40 & 1.27 \\
 2456517.78745 &  -1.66 & 1.23 \\
 2456518.81965 &  -1.70 & 1.25 \\
 2456534.80483 &   4.89 & 1.21 \\
 2456535.80156 &   0.75 & 1.28 \\
 2456539.79917 &   5.14 & 1.36 \\
 2456540.82230 &  -4.63 & 1.55 \\
 2456541.85331 &  -6.12 & 1.16 \\
 2456542.75722 &  -1.48 & 1.02 \\
 2456547.79837 & -11.14 & 1.29 \\
 2456548.77462 &  -8.03 & 1.21 \\
 2456562.80405 &  -3.92 & 1.05 \\
 2456563.71147 &  -5.93 & 1.08 \\
 2456569.79719 & -27.84 & 4.34 \\
 2456570.79627 &  -5.96 & 1.55 \\
 2456573.72675 &  -9.89 & 1.44 \\
 2456577.80143 & -17.01 & 1.00 \\
 2456581.79176 &  -6.52 & 0.78 \\
 2456582.73480 &  -2.58 & 0.63 \\
 2456583.69447 & -11.73 & 1.09 \\
 2456588.68050 & -12.69 & 0.57 \\
 2456589.77159 & -11.57 & 1.02 \\
 2456590.66789 & -11.63 & 1.32 \\
 2456591.66285 &  -9.16 & 0.74 \\
 2456596.59997 &  -3.76 & 1.12 \\
 2456597.69329 &  -0.91 & 0.72 \\
 2456606.66048 &  -1.10 & 0.84 \\
 2456607.69387 &  -8.66 & 1.00 \\
 2457192.94571 &  -9.75 & 1.23 \\
 2457225.99709 & -21.57 & 0.80 \\
 2457706.65722 &  43.99 & 1.23 \\
 2458239.99803 &  28.47 & 1.08 \\
 2458292.79881 &  17.20 & 1.81 \\
 2458384.63820 &  18.05 & 0.93 \\
 2458408.59895 &   8.73 & 0.79 \\
 2458409.59737 &   1.57 & 0.78 \\
 2458411.59738 &   9.05 & 0.90 \\
 2458413.59601 &   8.92 & 0.76 \\
 2458415.59282 &   9.33 & 0.80 \\
\hline
\end{tabular}
\end{table}

\section{Previous Analysis} \label{prev}
The first analysis of the planetary system around HD 200964 was given by \citetalias{jphc11}, using the ``Lick" and ``Keck11" datasets. These authors first perform a Markov chain Monte Carlo (MCMC) analysis of the system assuming Keplerian orbits for both of the planets in the system, i.e. neglecting planet-planet interactions. They use the results of this Keplerian MCMC to initialize a Differential Evolution Markov Chain Monte Carlo (DEMCMC) algorithm. The theoretical radial velocity at a given time is calculated using an $N$-body integrator, with a constraint that the system must remain stable for 100 years. They then perform rejection sampling on their final posterior, throwing out points which are not stable for $10^7$ years. Their best-fit, long term stable solution appears to have an RMS scatter of 28.1 m/s, which would indicate poor agreement between the model and the data. Furthermore, as also reported by \citet{tcm15}, we find that the best fit solution reported by \citetalias{jphc11} does not exhibit long-term stability, regardless of whether the reported orbital elements are taken to be astrocentric or Jacobi. However, \citetalias{jphc11} do not appear to specify the epoch at which the planets have the reported orbital elements. When planet-planet interactions are included, the orbital elements of the planets change as a function of time. Thus, in order to fully specify an orbit, the time at which the orbital elements are referenced must be stated in addition to the elements themselves. For example, for the parameters given by \citetalias{jphc11}, the period of the outer planet ranges from $\sim 772$ to $857$ days over the timescale of the radial velocity observations. Given the degree to which the orbital elements change over the timescale of the observations for the parameters reported by \citetalias{jphc11}, it is quite possible that the discrepancy we find between their best-fit solution and the data is because the epoch to which the elements are referenced is not specified.

The reported 4:3 MMR exhibited by the system is interesting, as it is quite difficult to capture planets of gas giant mass into this resonance through convergent migration alone, as discussed by \cite{rpv12}. Subsequent works have explored the stable regions of parameter space for the parameters reported by \citetalias{jphc11}, (\citealt{wht12}), investigated in more detail the resonant behavior exhibited for the reported parameters (\citealt{mk_2016}) and investigated other, more complex scenarios for the formation of HD 200964 (\citealt{e_2012}, \citealt{tcm15})

\section{Methods} \label{methods}
In this work, in addition to analyzing a baseline of data longer than that used in \citetalias{jphc11}, we investigate the underlying posterior by explicitly conditioning our Markov chain Monte Carlo (MCMC) search on long-term stability. MCMC is a commonly used method to sample from a probability distribution (see e.g. \citealt{s_2017} for a review); in this context it used to sample from the posterior probability distributions for the orbital parameters of the planetary system (as well as the stellar jitter, see below). In this section we specify the methods employed to find these stable, best-fit solutions. We begin by investigating best-fit solutions including planet-planet interactions but neglecting stability (Section \ref{fitting}). After constructing the posterior distribution of orbital parameters without stability, we show that ``rejection sampling'', i.e. discarding solutions that do not exhibit long-term stability, yields few long-term stable solutions (Section \ref{rej}). Thus, to improve our measurement of the long-term stable posterior distribution, we explore parameter space using a likelihood function that explicitly takes stability into account (Section \ref{stab_fits}). We find that this method does a much better job of fitting the posterior distribution, though we find the posterior is multi-modal (Section \ref{post}). Finally, we perform a Monte Carlo search to verify after the fact that we have identified all relevant stable regions of parameter space (Section \ref{dart}).

\subsection{Fits Incorporating Planet-Planet Interactions} \label{fitting}
\begin{figure}[htbp]
	\centering
	\plotone{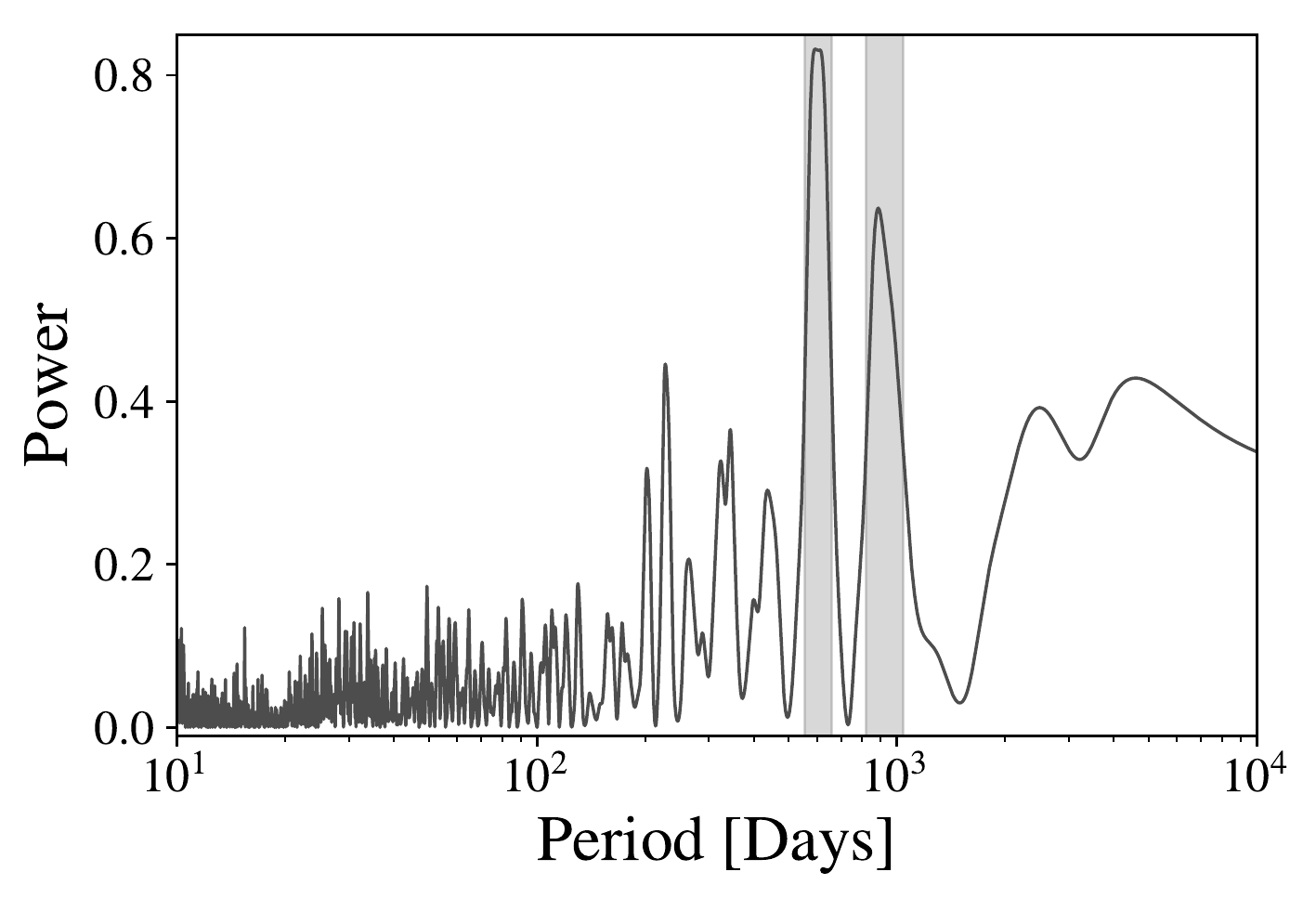}
		\caption{A generalized Lomb-Scargle periodogram for the RV data of HD 200964. Note the two strong peaks at $\sim$600 and $\sim$900 days, demonstrating that the system likely features two closely-packed planets. The full width at half maximum of each peak is indicated by the gray rectangle.}
\label{fig:GLS}
\end{figure}

We begin our analysis by searching for fits to the RV data without explicitly requiring our solutions to be stable. Firstly, we note that inspection of a usual generalized Lomb-Scargle periodogram (GLS, see e.g. \citealt{zk_2009}), leads inexorably to the conclusion that the two planets in the system are closely packed. A GLS for the RV data shown in Figure \ref{fig:rvt_nostab} is plotted in Figure \ref{fig:GLS}. The two largest peaks (note that we have omitted a peak at $\sim$1 day which is likely an alias of the sampling period of the data) of the GLS are near $\sim$600 and $\sim$900 days. While the actual periods of the planets we determine will be affected by planetary eccentricity and dynamical interaction between the planets, these close peaks nonetheless indicate that the system likely contains two closely packed planets.

\begin{figure} [h]
	\centering
	\includegraphics[width=1.15\linewidth]{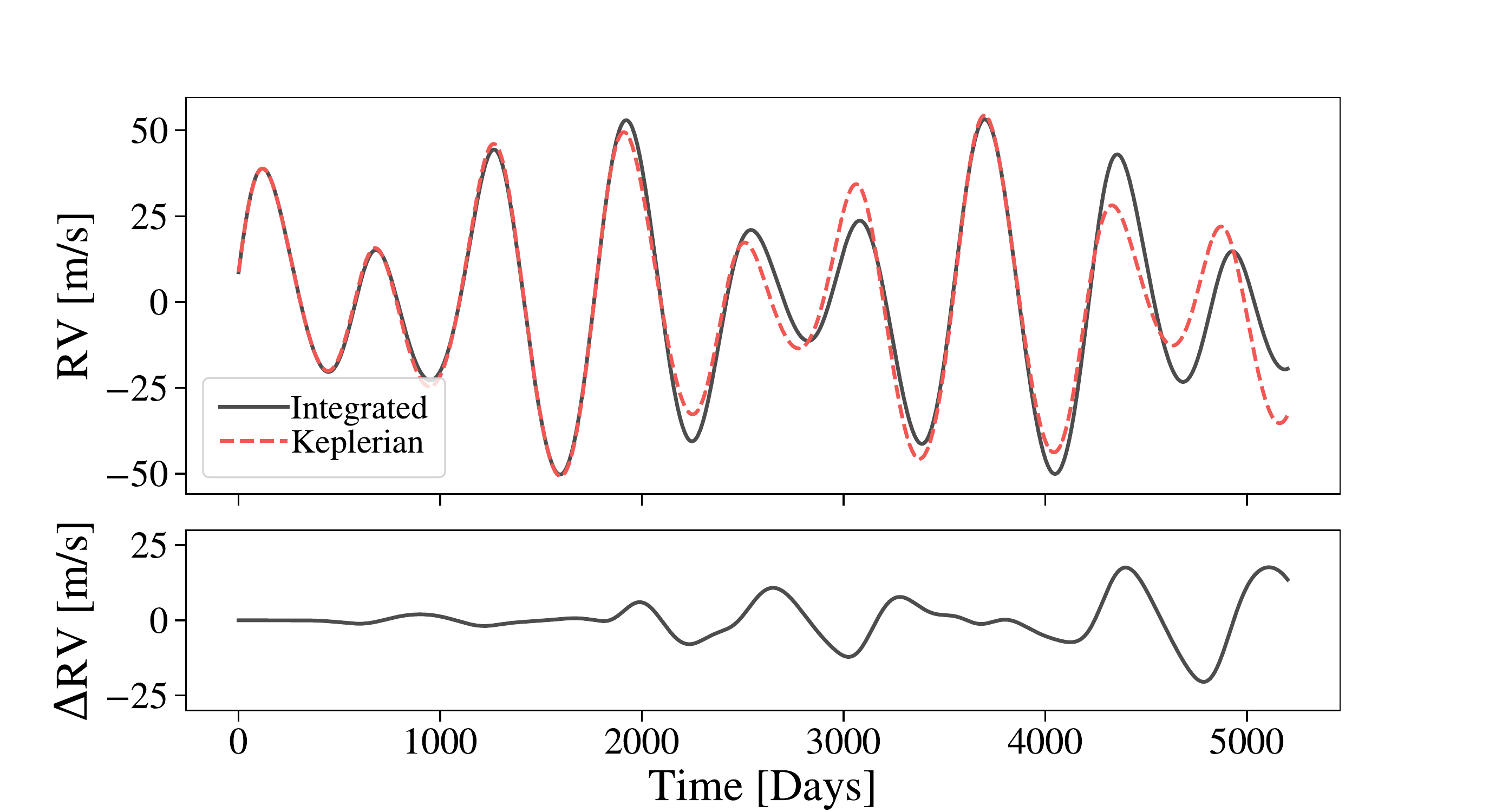}
	\caption{A comparison of the radial velocity determined by numerically integrating the motions of the planets and by advancing the planets forward on Keplerian orbits. The orbital parameters used are our best-fit long-term stable solution, as discussed in Section \ref{stab_fits}. The top panel shows the stellar radial velocity determined by the two methods, while the bottom shows the difference in the two curves. There is substantial disagreement between the integrated and Keplerian radial velocities due to the strong planet-planet interactions present.}
	\label{fig:int_kep_comp}
\end{figure}

Thus the gravitational interactions between the planets constitute an important component to the observed radial velocity of HD 200964, and cannot be neglected. Often, theoretical radial velocity values are calculated by advancing the planets along Keplerian orbits, in effect neglecting any perturbations between the planets. For non-closely packed systems this is generally a fine approximation, as perturbations between the planets are unimportant over the timescale of the RV observations. As illustrated in Figure \ref{fig:int_kep_comp}, however, this is not the case for HD 200964. Figure \ref{fig:int_kep_comp} plots the radial velocity as a function of time determined by both using only Keplerian orbits, as well as a full $N$-body integration of the equations of motion. The difference between the two values is shown in the bottom panel. The orbital parameters used correspond to our best-fit, long-term stable solution (see Section \ref{stab_fits}). 

Clearly, neglecting the planet-planet interactions is a poor approximation; in what follows all calculation of radial velocity values will be done by numerically integrating the star-planet system forward in time. In order to perform our numeric integrations, both to calculate the theoretical radial velocity and to determine the lifetime of our planetary systems (see Sections \ref{rej} and \ref{stab_fits}), we use the $N$-body integration package \texttt{REBOUND} (\citealt{rl12}).

For the purpose of computing the theoretical radial velocity for comparison with the observations,  we use the IAS15 integrator (\citealt{rs15}), which is a 15th order integrator with adaptive time stepping. All orbital elements provided in this paper are quoted relative to the primary star, i.e. they are astrocentric coordinates, and are given at the epoch of the first data point, i.e. JD 2453213.895. Following \citet{bfv_2016}, we take the central star to have mass $M_* = 1.45 M_\odot$. We use a usual radial velocity coordinate system, such that the inclination $i$ represents the angle between orbital plane and the plane of the sky, which we take to be the reference plane. The argument of periapse, $\omega$, is the angle between the line of ascending nodes and the periapse direction. The observer is taken to lie in the $-\hat{z}$ direction relative to the reference plane; in keeping with convention velocities in this direction, i.e. towards the observer, are quoted as positive. For clarity, due to the strong planet-planet interactions we specify the mean longitudes of the two planets at epoch, $\lambda$, as opposed to the planets' time of periastron passage. In this work we fix $i=90^\circ$, corresponding to edge on orbits, and fix the longitude of ascending node, $\Omega=0$. We comment on the degeneracy between the system's inclination and the masses of the planets in Section \ref{post}.

\begin{figure}[htbp]
	\centering
	\includegraphics[trim=70 0 0 20, clip, width=1.1\linewidth]{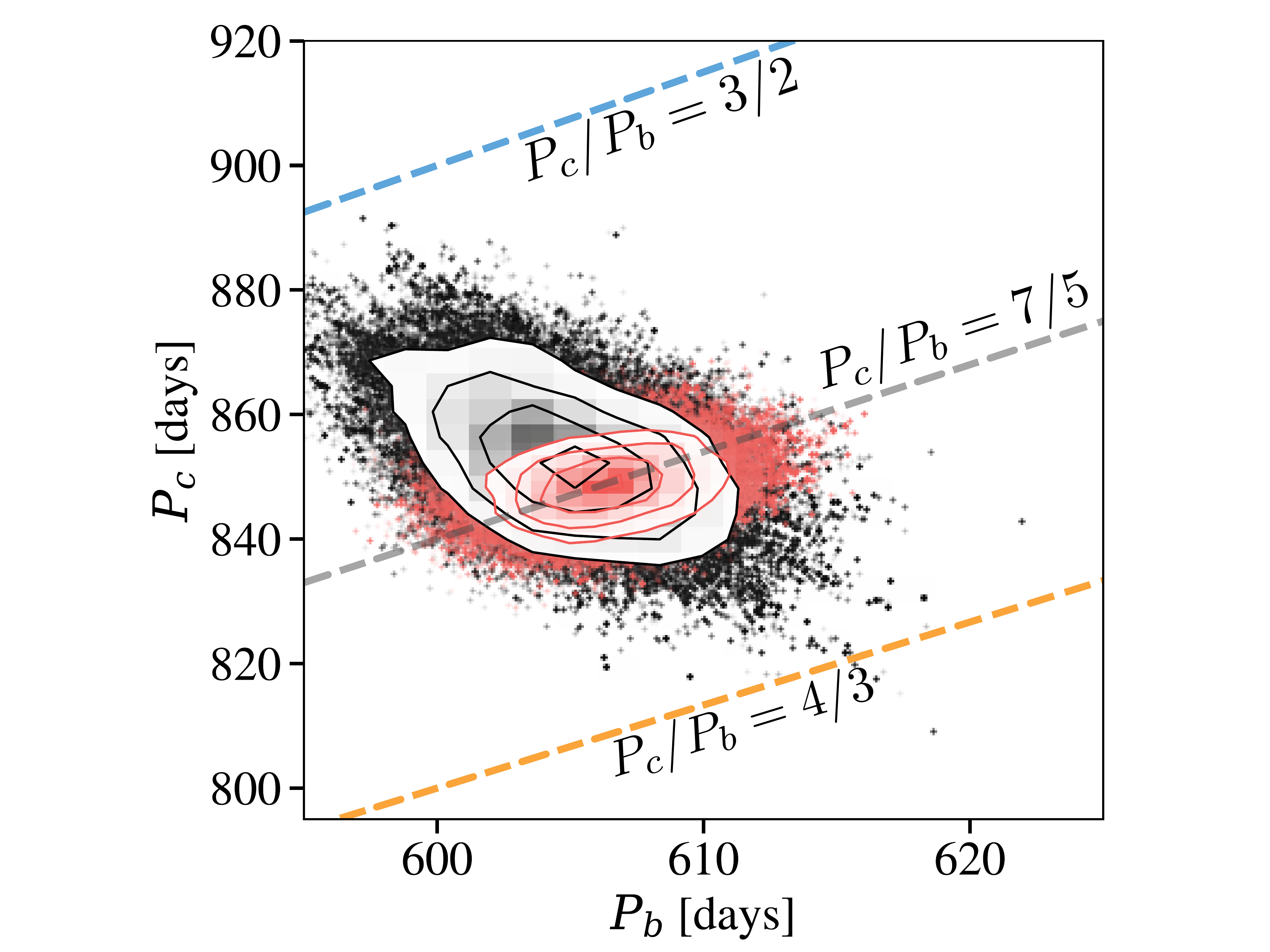}
		\caption{2D histograms of the posterior distributions for the planets' periods, using $N$-body integration to calculate the radial velocity but without long-term stability (black points, see Section \ref{fitting}) and advancing the planets on Keplerian orbits (red points). Lines denoting exact ratios of $P_c/P_b$ are shown for ratios of 3:2 (blue), 7:5 (gray) and 4:3 (orange).}
\label{fig:per_ratio_comp}
\end{figure}

Following other works (e.g \citealt{jfm_2007}, \citealt{cbm_2008}, \citetalias{jphc11}), we introduce a ``stellar jitter'' term in our fitting, which is an additional error term that is added in quadrature to the ``known'' error, i.e. the error on each measurement is taken to be $\sqrt{\sigma_k^2 + \sigma_j^2}$, where $\sigma_k$ is the given error and $\sigma_j$ is the proposed value of the stellar jitter term. We also note that we are using a single value to characterize the stellar jitter, meaning that we are neglecting variation in jitter between different instruments (\citealt{b_2009}). We have checked that the inclusion of multiple jitters has no qualitative effect on the posterior distribution shown in Figure \ref{fig:per_ratio_comp}. However, fitting a different jitter for each data set (as is done in e.g. \citealt{nrp_2016} or \citealt{mlt_2018}) would allow us to characterize the difference in instrumental noise between the various datasets.

We calculate the likelihood for a given set of orbital parameters by assuming that the radial velocity measurements are all independent and Gaussian distributed, with error given by $\sigma_i = \sqrt{\sigma_k^2 + \sigma_j^2}$, as discussed in the preceding paragraph. In this case, the log likelihood $\mathcal{L}$ is given by
\begin{align}
    \mathcal{L} = -\sum_i \left[ \frac{\left(v_i - RV(t_i) - O_D\right)^2}{2 \sigma_i^2} + \log\left(\sigma_i \sqrt{2 \pi} \right) \right]
\end{align}
where $v_i$ are the measured radial velocities and $RV(t_i)$ are the model radial velocities. Here $O_D$ refers to the constant offset to each dataset (see Section \ref{obs}), which must also be fit, introducing 4 additional parameters into our fitting. Instead of including the 4 offsets as parameters in our MCMC search, the offsets are separately optimized for every proposed set of orbital parameters. That is, once the model radial velocities are known, it is straightforward to show that the constant offset to each dataset that maximizes the likelihood can be obtained by calculating the weighted mean of the difference between the model and the data
\begin{align}
    O_D = \frac{1}{S} \sum_{i \in  D} \frac{v_i - RV(t_i)}{\sigma_i^2} 
\end{align}
where $S \equiv \sum_i 1/\sigma_i^2$. This simplifies our fitting algorithm, but does mean that we may miss degeneracies between the constant offsets and the orbital elements. 

For our priors, we assume uniform probability in some specified domain for each parameter, except for the planetary eccentricities, where the priors are uniform in log space. For periods of each planet, the priors are uniform between 400 to 1000 days for planet b, and 500 to 1100 days for planet c. The prior on plantary mass is uniform between 0.1 and 10 $M_J$ for both planets. For the planetary eccentricity, the prior is uniform in log space between -4.5 and 0. For all the angles, the priors are taken to be uniform between $-720^\circ$ and $720^\circ$. This is done to ensure that the arguments of pericenter do not diverge to arbitrarily large values when the planet's eccentricity is low. In practice the actual values of parameters in our searches are quite far from the limiting bounds, with the exception of planetary eccentricity and the corresponding argument of pericenter, where the bounds are important for cases of low eccentricity.

To explore the parameter space, we initially use the \texttt{scipy} minimizer to optimize the orbital parameters. We initially fix the orbital periods and masses of the planets, using the GLS and the amplitude of the RV signal to provide rough estimates of these parameters, and perform an optimization on the rest of the parameters, starting from random values. We chose five of these optimizations which both had high likelihood and different final parameters to initialize our MCMCs.

We used the software \texttt{emcee} \citep{fhl13} to perform our MCMC search. We initialized different MCMC searches from our converged optimizations. We let these MCMCs run for $\sim$1000 steps, and look at the regions of high likelihood. We found that all of these searches identify a single region as having the highest likelihood. We then reinitialized a final search in this region. We ran this MCMC for an initial burn in period, then discarded these walker positions and ran the MCMC to convergence. To asses convergence of our MCMC runs, we used the potential scale reduction factor (PSRF, \citealt{gr_1992}). A common method to asses convergence is to run the MCMC until the PSRF for every parameter has a value $<1.1$ \citep{bg_1998}. However, for our MCMC runs the PSRF for the two eccentricities and arguments of pericenter often do not fall below 1.1, likely because at low eccentricities the posterior probability is completely insensitive to these parameters. Thus, in practice we consider our MCMC converged if the PSRF for all parameters, except for the two eccentricities and two arguments of pericenter, is below 1.1. 

A corner plot showing our best fit posterior distribution for the orbital parameters is shown in Appendix \ref{corner_plots} (Figure \ref{fig:nostab_corner}). The model radial velocity produced from our best-fit parameters (maximum likelihood) is shown in Figure \ref{fig:rvt_nostab}, the median values of our posterior distribution are given in Table \ref{tab:median_nostab}, and the maximum likelhiood orbital parameters are given in Table \ref{tab:bestfit_nostab}.

\begin{figure}[htbp]
	\centering
	\includegraphics[trim=50 0 0 40, clip, width=1.1\linewidth]{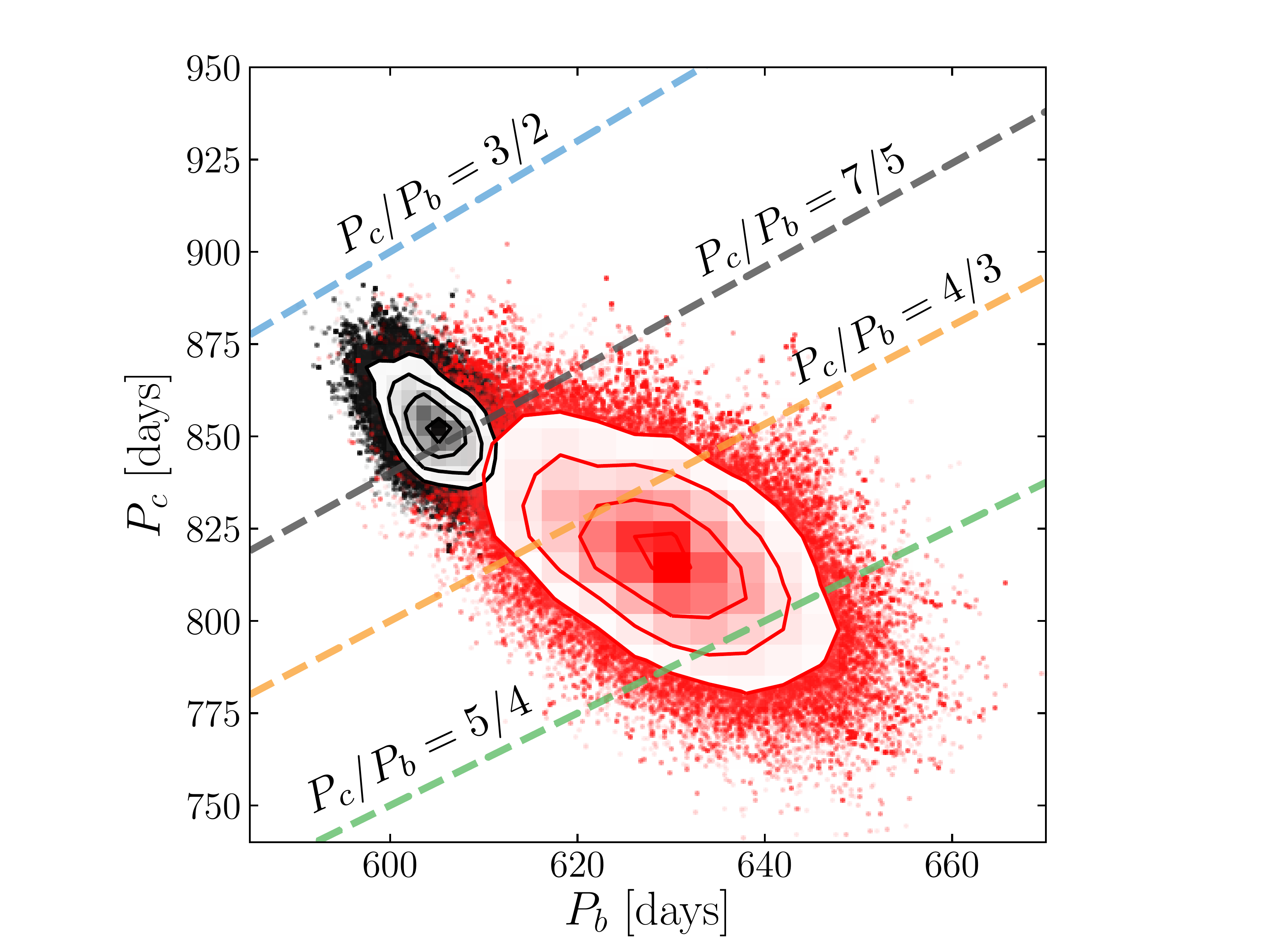}
		\caption{2D histograms of the posterior distributions for the planets' periods, using $N$-body integration to calculate the radial velocity but without long-term stability. The black points show the posterior produced by using the full dataset, while the red points show the posterior obtained by analyzing only the \citetalias{jphc11} data. Lines denoting exact ratios of $P_c/P_b$ are shown for ratios of 3:2 (blue), 7:5 (gray), 4:3 (orange), and 5:4 (green).}
\label{fig:per_ratio_comp_early}
\end{figure}

The periods of the planets in our posterior distribution are much more constrained than the results obtained by \citetalias{jphc11}. The median period ratio of the system has also moved to $P_c/P_b \sim 7/5$, whereas \citetalias{jphc11} found values much closer to $4/3$. This is due to our observations spanning a longer timescale. To illustrate this point, in Figure \ref{fig:per_ratio_comp_early} we plot the posterior for our new data along with the $N$-body integrated posterior distribution produced by analyzing just the \citetalias{jphc11} data (see Section \ref{reanaly}). This is consistent with the results of \citet{lbw_2019}, who also report the period of planet c to be around 850 days based on a Keplerian fit to the data.

Interestingly, using $N$-body integration to determine the theoretical RV values broadens the posterior distribution of $P_b$ and $P_c$ compared to a purely Keplerian fit for the full dataset. For comparison with our $N$-body integrated fits, we repeat our analysis with the assumption of Keplerian orbits for both planets. The 2D histogram of a Keplerian fit to the data is plotted in red in Figure \ref{fig:per_ratio_comp}. In particular, it appears that the dynamical interaction between the planets allows for period ratios close to both 3:2 and 4:3 to fit the data, which are more strongly ruled out in a purely Keplerian fit. 

\begin{center}
\begin{deluxetable*}{ccc}
	\tabletypesize{\footnotesize}
	\tablecaption{Median orbital parameters, no long term stability}
	\tablehead{\colhead{Parameter\tablenotemark{a}}& \colhead{HD 200964 b}& \colhead{HD 200964 c}}
	\startdata
	Orbital Period,  $P$ [days], & $604.69^{+3.38}_{-3.10}$ & $852.55^{+9.42}_{-8.30}$ \\
	Mass, $m$ $\left[M_J\right]$  & $1.72^{+0.05}_{-0.05}$ & $1.20^{+0.06}_{-0.06}$ \\
	Mean longitude, $\lambda$ [deg] & $307.40^{+5.26}_{-5.06}$ & $239.47^{+6.27}_{-6.42}$\\
	Argument of periastron, $\omega$ [deg] & $294.48^{+21.08}_{-22.70}$ & $259.32^{+57.71}_{-47.07}$\\
	$\log_{10}$ Eccentricity, $e$ & $-1.15^{+0.11}_{-0.15}$ & $-1.49^{+0.55}_{-1.81}$\\
	Stellar Jitter, $\sigma_j$ [m/s] & $6.05^{+0.46}_{-0.39}$ &
    \enddata
    \vspace{1mm}
	\tablenotetext{a}{Values for orbital elements are in astrocentric coordinates, are referenced to the epoch of the first data point, JD 2453213.895, and assume an inclination $i=90^\circ$. The reported values are median values for the posterior distribution, and the reported error bars are 84\% and 16\% quantiles.} 
	\label{tab:median_nostab}
\end{deluxetable*}

\begin{deluxetable*}{ccc}
	\tabletypesize{\footnotesize}
	\tablecaption{Maximum likelihood orbital parameters, no long term stability}
	\tablehead{\colhead{Parameter\tablenotemark{a}}& \colhead{HD 200964 b}& \colhead{HD 200964 c}}
	\startdata
	Orbital Period,  $P$ [days], & 607.7 & 845.3 \\
	Mass, $m$ $\left[M_J\right]$  & 1.71 & 1.21 \\
	Mean longitude, $\lambda$ [deg] & 312.5 & 233.7\\
	Argument of periastron, $\omega$ [deg] & 297.4 & 270.5\\
	$\log_{10}$ Eccentricity, $e$ & -1.13 & -0.92\\
	Stellar Jitter, $\sigma_j$ [m/s] & 5.60 &
    \enddata
    \vspace{1mm}
	\tablenotetext{a}{Values for orbital elements are in astrocentric coordinates, are referenced to the epoch of the first data point, JD 2453213.895, and assume an inclination $i=90^\circ$.} 
	\label{tab:bestfit_nostab}
\end{deluxetable*}
\end{center}

\begin{figure}[htbp]
	\centering
	\includegraphics[width=\linewidth]{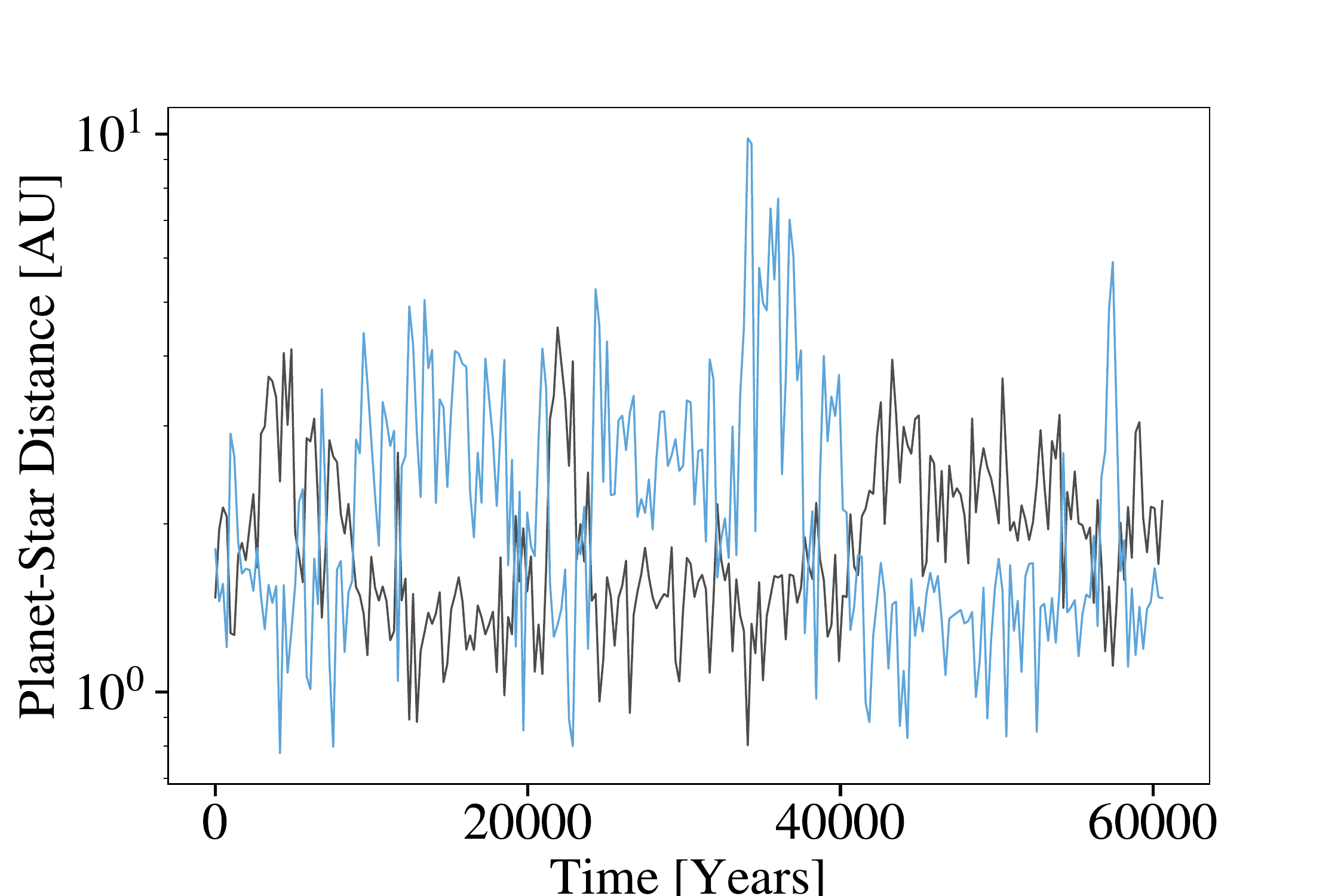}
        \epsscale{2.0}
		\caption{Distance between planet b (black line) and planet c (blue line), and the host star for our best-fit solution without long-term stability (see Section \ref{fitting}). The planets experience large, non-periodic fluctuations in distance from the star due to their strong mutual perturbations. The short timescale of these fluctuations relative to the age of the host star makes it unlikely, if the proposed best fit solution were correct, that the system would be observed in the original orbital configuration. Furthermore, these fluctuations are a strong indication that the system will become unstable on timescales much less than the age of the system. This is indeed the case---planet c is eventually scattered to a distance $> 100\, \rm{AU}$ on $10^5$ year timescales.}
\label{fig:d_planet_star}
\end{figure}

Closer examination of our $N$-body integrated posterior distribution shows that many of the points, including our best fit solution, feature extremely close encounters between the two planets. An example from our best-fit parameters is shown in Figure \ref{fig:d_planet_star}, which plots the distance between each planet and the central star as a function of time. While neither of the planets is ejected over this timescale, the two planets, particularly the outer planet, experience large amplitude fluctuations in distance from the central star. Thus, it is extremely unlikely, if the system were truly in this orbital configuration initially, that we would observe it before the configuration changed substantially. Furthermore, integration over long time scales indicates the outer planet is scattered out past 100 AU on $10^5$ year timescales.

The majority of the solutions in Figure \ref{fig:per_ratio_comp} do not exhibit long term stability. 

\subsection{Rejection Sampling} \label{rej}
In order to find long term stable solutions, we begin by using ``rejection sampling" on the posterior distribution found in Section \ref{fitting}. Rejection sampling is less computationally intensive than doing a full search conditioned on stability, and has been employed in other works to find best fit orbital parameters for planetary systems which are also stable (e.g. \citealt{wgd_2018}). In rejection sampling, we first construct a posterior distribution for the planetary system that does not take stability into account. Some fraction of the points (or, in our case, all of the points) in the posterior are chosen at random, and are then tested for long-term stability. All of the points in the posterior that pass the stability criteria then make up the new best-fit posterior which is conditioned on stability. 

The converged posterior distribution shown in Figure \ref{fig:per_ratio_comp} contains 287,296 points in parameter space. We then tested all of these points for stability for $10^3$ orbital periods of planet c. We consider systems stable if both planets remained between 150\% of their initial periastron distance and 50\% of their apastron distance from the central star during the course of the integration. We considered distance from the central star, as opposed to the semi-major axis of the planets, as many of our best fit solutions feature extremely close encounters between the planets, as discussed above. This can cause the semi-major axis of planet b to diverge as its velocity is temporarily excited to above the escape velocity from the system, despite the fact that the system remains stable after this close encounter. Though it is unlikely a system featuring such a close encounter will survive on long timescales, we did not want to prematurely discard these solutions without checking for long term stability. These integrations were again carried out using the IAS15 integrator.

Of the points in the initial posterior, 2,295, i.e. $<1\%$ of the systems survived for $10^3$ orbital periods. We then tested these remaining points for longer term stability: each set of orbital parameters was integrated for $10^7$ orbital periods of planet c. For these long term stability analyses we use the WHFAST integrator (\citealt{rt15}), an implementation of the sympletic Wisdom-Holman integrator. 
Unless otherwise noted, we set the timestep for our integrations with WHFAST to be $dt = P_{\rm{min}}/100$, where $P_{\rm{min}}$ is the shortest initial orbital period of the planets in the system. This is five times shorter than the orbital period recommended by \cite{dll98}, who recommend $dt = P_{\rm{min}}/20$ for a second order sympletic integrator. Of the points tested, only 1,111 survive for $10^7$ orbital periods. This is far too few points to construct a converged posterior for stable, best fit solutions to the data. We would require 1-2 orders of magnitude more points in our original, non-stable posterior, in order to retain enough points in the rejection sampling to construct a converged stable posterior, which would be extremely computationally intensive. Though the posterior obtained through rejection sampling is clearly not converged, the points do appear to lie in the general region of parameter space identified in Section \ref{stab_fits}. With rejection sampling, we only identify stable fits near 7:5 period ratio (c.f. Figure \ref{fig:per_ratio_comp_stab}, purple points), while the broader search described in Section \ref{stab_fits} identifies other possible period ratios. 

It is also interesting to note that when this exercise was carried out for fits on just the Keck and APF datasets (i.e. omitting the Lick and Keck11 datasets), \textit{none} of the points in the initial posterior survived for $10^7$ orbital periods. It is only when we have data spanning a longer timescale that we appear to be able to find \textit{any} best-fit solutions that also exhibit stability. We suspect that this effect stems from the longer time baseline and better coverage of the RV signal that inclusion of the two later data sets provides. As more data is included the parameters of the planets in the system become better constrained, and our posterior distribution moves closer to the ``true" parameters of the underlying system, which presumably does exhibit long term stability. Thus, with more data, we expect a greater likelihood that the posterior distribution we construct without explicitly including stability will overlap with stable regions of parameter space.  

While rejection sampling is insufficient to construct a converged posterior distribution, some of these points are useful places to initialize MCMC searches with stability included, which we discuss in the next section.

\subsection{Likelihood Function Conditioned on Stability} \label{stab_fits}

As rejection sampling is insufficient to produce a converged posterior distribution, we therefore try a different approach---we modify the likelihood function used in performing searches of parameter space by setting the likelihood function to be 0 if the system is not found to be stable for a predetermined period of time.  We consider a planetary system to be stable if the semi-major axes of both planets remain between 50\% and 150\% of their initial values. This means that any samples in our final posterior distributions now exhibit long-term stability, but also means that our search has trouble exploring between stable regions of parameter space. If we were merely looking for maximum likelihood solutions, the stable solutions we found through rejection sampling would be sufficient for initializing our long-term stable MCMC searches. However, given the large upwards shift in period ratio that occurs when more data is included in the fitting when compared to the \citetalias{jphc11} data (see Figure \ref{fig:per_ratio_comp_early}), we feel it is quite important to explore other possible modes near the best-fitting solutions, since it is quite possible, as we discuss below, that additional frequencies introduced by the dynamical interaction between the planets are obscuring the underlying period ratio. Due to this difficulty, we use several different methods and initializations for our search, which we discuss in detail below. We ultimately identify three peaks in our posterior distribution, which are discussed in Section \ref{post}. We do require multiple different initializations to find these various modes, which leaves open the question of whether other initialization methods might find additional modes in the posterior distribution. We return to this question in Section \ref{dart}.

The simplest method of initialization, as well as the method that overall finds the best-fitting region of parameter space, is to simply initialize our MCMC near the best-fit solution found by rejection sampling. This method produces a peak near a 7:5 period commensurability, which is unsurprising given that this is where our non-stable posterior distribution is located. 
For another initialization, we use a genetic algorithm (GA) to explore the parameter space. As we suspect there may be multiple local maxima of our posterior distribution, a GA may be useful to identify these different maxima and ultimately identify the global maximum. We use the open-source optimization framework Pyevolve \citep{p_2009}. Our genetic algorithm calculates likelihood scores using the same criteria discussed above, i.e. log-likelihood derived from assuming the observations are Gaussian distributed and independent, conditioned on long-term stability. The negative of the log-likelhiood is used as the ``fitness" for the GA. For our GA runs we test for stability for $10^6$ periods of planet c's orbit. We find that allowing the algorithm to evolve until an average fitness score of at least 800 is reached, or until there is no significant increase in likelihood between sequential generations, is sufficient time for the algorithm to find useful starting points for the MCMC. We initialize the MCMC in a small Gaussian ball around the best fit parameters determined by the genetic algorithm, and allow the MCMC to run to convergence. 

The GA strongly favors a region of stability similar to the parameters identified in Figure \ref{fig:nostab_corner}, but with the period of the larger planet closer to $\sim 900$ days, which places the system firmly in a 3:2 MMR, as discussed in Section \ref{res}. This region is extremely stable, making it easier for the GA to explore. The GA misses the stable region of parameter space near $P_c/P_b \sim 7/5$ identified by our rejection sampling in Section \ref{rej}; this is likely because the search by the GA is \textit{too} broad for this application, and the stable region near 7:5 is much more narrow than the region near 3:2. 

We also begin a search starting from the orbital parameters identified by \cite{tcm15}, who explored the formation and evolution of HD 200964 using the data of \citetalias{jphc11}, with a higher stellar mass of $M_* = 1.57 M_\odot$, and gave long-term stable solutions in the 4:3 MMR. The specific parameters reported in this work do not match the data well according to our model, likely because of a disagreement between the coordinate systems used. Thus, beginning with their reported planetary masses (scaled by a factor $M_p/M_*$) and eccentricities, we first optimize over angular parameters, before performing an optimization over all parameters and a subsequent MCMC search. This search does find stable solutions near a 4:3 period ratio that fit the data well, but the search also finds a smaller number of solutions near the 7:5. Though the walkers in our search spend more time near 4:3, solutions near 7:5 clearly have better posterior probability; it is likely the MCMC has difficulty moving between the two period ratios due to a dearth of stable solutions at period ratios intermediate between the two regions. We therefore initialize another MCMC at our best fit solution from the previous run. This MCMC converges to a region similar to the region identified by starting at the best-fit obtained through rejection sampling. 

Thus, we have identified three peaks in our posterior distribution---one near a 3:2 period ratio, another near a 4:3 period ratio, and peak containing our best fit solution near a 7:5 period ratio. In the next section we discuss these peaks in more detail.

\subsection{Final Posterior Distribution} \label{post}

We give median values of the orbital parameters from each mode of the posterior distribution in Table \ref{tab:median_stab}, and maximum likelihood parameters in Table \ref{tab:bestfit_stab}. Since the 4:3 distribution joins on to the 7:5 distribution, we remove all points with $P_c > 7/5 \, P_b$ before calculating the median or the errors. Theoretical radial velocity curves for the maximum likelihood parameters are shown in Figure \ref{fig:rvt_comp}. The full posterior distributions are plotted in Appendix \ref{corner_plots}. We also stress that it is more meaningful to talk about overall stable regions of parameter space rather than particular orbital configurations. Long-term orbital integrations are inherently chaotic, and lifetimes of a given set of orbital parameters can vary by an order of magnitude depending on the machine used to carry out the integration. 

All of our parameters discussed above are reported for $i=90^\circ$. Though there are still strong degeneracies between $M_p$ and $i$ in our modeling, we note both the theoretical RV signal and the long-term stability of the system are directly sensitive to the planetary mass $M_p$, as opposed to just $M_p \sin i$, which is the relevant quantity when planets are allowed to move on purely Keplerian orbits. One extension of our work would be to directly constrain the masses of the planets by allowing the overall inclination of the system to vary, while still keeping the planets coplanar. We could also allow mutual inclinations between the planets, which would necessitate allowing $\Omega$ to vary. This could improve our stability constraints, and allow us to further constrain $M_p$. We leave these investigations as avenues for future work.

We also note that all three posterior distributions identified, that is, near period ratios of 3:2, 4:3, and 7:5, feature a long tail in the eccentricity of planet c consistent with planet c on a circular orbit. We therefore re-run our MCMC, now setting planet c to be circular, which eliminates two parameters from our fitting. The resultant searches identify very similar regions of parameter space to the solutions with eccentricity included, but none of the solutions are truly consistent with planet c being circular. Instead, planet c's eccentricity is quickly excited by the companion, and, over longer timescales, both planets' eccentricities oscillate, with average values that are both of order $10^{-1}$. We also comment that for two planets to be in MMR, the ``test" particle must have some eccentricity. Thus, in what follows we use our orbital solutions with eccentricity included. 
\begin{center}
\begin{deluxetable*}{c|cc|cc|cc} 
	\tabletypesize{\footnotesize}
	\tablecaption{Median orbital parameters, $10^6 P_c$ stability}
	\tablehead{\colhead{Parameter\tablenotemark{a}}\vrule& \colhead{HD 200964 b, 7:5}& \colhead{HD 200964 c, 7:5} \vrule& \colhead{HD 200964 b, 4:3}& \colhead{HD 200964 c, 4:3} \vrule& \colhead{HD 200964 b, 3:2}& \colhead{HD 200964 c, 3:2}}
	\startdata
	Orbital Period,  $P$ [days] & $603.27^{+2.33}_{-2.17}$ & $854.46^{+4.56}_{-4.39}$ & $605.85^{+2.53}_{-2.48}$ & $837.51^{+4.62}_{-6.12}$ & $598.70^{+2.79}_{-2.77}$ & $881.11^{+7.62}_{-6.62}$ \\
	Mass, $m$ $\left[M_J\right]$  & $1.72^{+0.05}_{-0.05}$ & $1.16^{+0.05}_{-0.05}$ & $1.74^{+0.05}_{-0.05}$ & $1.13^{+0.05}_{-0.06}$ & $1.68^{+0.06}_{-0.06}$ & $1.26^{+0.07}_{-0.07}$ \\
	Mean longitude, $\lambda$ [deg] & $307.90^{+4.32}_{-4.04}$ & $236.76^{+4.28}_{-4.50}$ & $311.31^{+4.49}_{-4.46}$ & $223.98^{+4.70}_{-4.91}$ & $287.17^{+6.15}_{-4.57}$ & $269.31^{+5.52}_{-5.80}$\\
	Argument of periastron, $\omega$ [deg] & $325.762^{+13.16}_{-13.51}$ & $252.58^{+112.94}_{-103.12}$ & $293.97^{+14.15}_{-13.89}$ & $273.05^{+96.72}_{-118.05}$ & $317.12^{+17.78}_{-19.11}$ & $169.28^{+160.12}_{-35.34}$\\
	$\log_{10}$ Eccentricity, $e$ & $-1.21^{+0.05}_{-0.05}$ & $-3.10^{+0.90}_{-0.98}$ & $-1.16^{+0.06}_{-0.05}$ & $-2.99^{+1.02}_{-1.06}$ & $-1.12^{+0.14}_{-0.19}$ & $-1.47^{+0.38}_{-1.91}$\\
	Stellar Jitter, $\sigma_j$ [m/s] & $6.27^{+0.42}_{-0.40}$ & \, & $6.57^{+0.47}_{-0.42}$ & \, & $7.47^{+0.53}_{-0.49}$ & \,
    \enddata
    \tablenotetext{a}{Values for orbital elements are in astrocentric coordinates, are referenced to the epoch of the first data point, JD 2453213.895, and assume an inclination $i=90^\circ$. The reported values are median values for the posterior distribution, and the reported error bars are 84\% and 16\% quantiles.}
 \label{tab:median_stab}
\end{deluxetable*} 
\begin{deluxetable*}{c|cc|cc|cc} 
	\tabletypesize{\footnotesize}
	\tablecaption{Maximum Likelihood orbital parameters, $10^6 P_c$ stability}
	\tablehead{\colhead{Parameter\tablenotemark{a}}\vrule& \colhead{HD 200964 b, 7:5}& \colhead{HD 200964 c, 7:5} \vrule& \colhead{HD 200964 b, 4:3}& \colhead{HD 200964 c, 4:3} \vrule& \colhead{HD 200964 b, 3:2}& \colhead{HD 200964 c, 3:2}}
	\startdata
	Orbital Period,  $P$ [days] & 601.5 & 856.8 & 605.6 & 839.3 & 598.8 & 886.4 \\
	Mass, $m$ $\left[M_J\right]$  & 1.75 & 1.18 & 1.77 & 1.16 & 1.72 & 1.33 \\
	Mean longitude, $\lambda$ [deg] & 304.7 & 238.5 & 308.1 & 227.6 & 286.4 & 272.8\\
	Argument of periastron, $\omega$ [deg] & 327.1 & 246.2 & 304.1 & 293.8 & 304.1 & 181.1\\
	$\log_{10}$ Eccentricity, $e$ & -1.18 & -2.02 & -1.3 & -3.36 & -1.12 & -1.08\\
	Stellar Jitter, $\sigma_j$ [m/s] & 6.1 & \, & 6.4 & \, & 7.2 & \,
    \enddata
    \tablenotetext{a}{Values for orbital elements are in astrocentric coordinates, are referenced to the epoch of the first data point, JD 2453213.895, and assume an inclination $i=90^\circ$.}
 \label{tab:bestfit_stab}
\end{deluxetable*} 
\end{center}
\begin{figure*}[htbp]
	\centering
\includegraphics[width=7in]{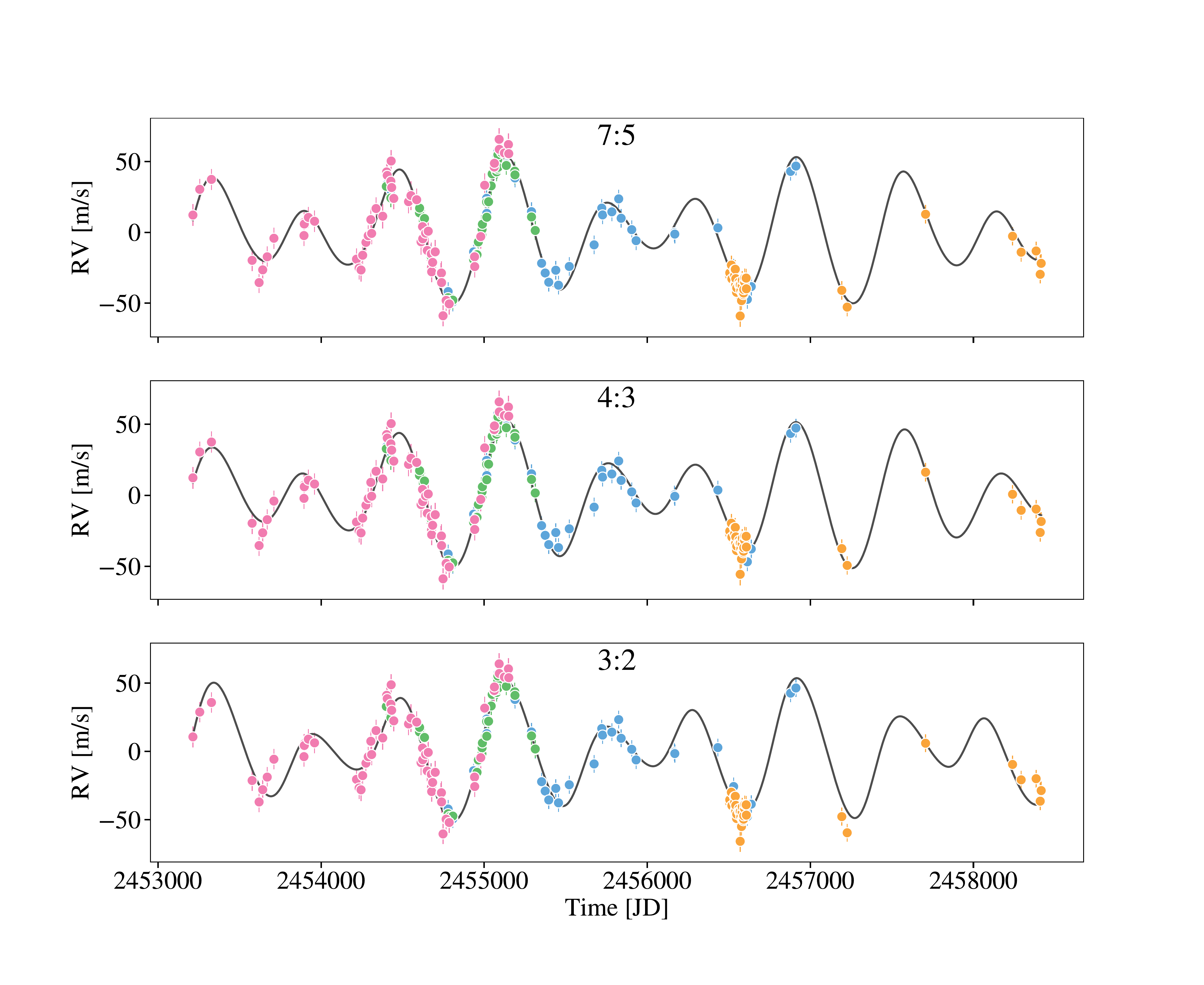}
		\caption{Comparison of the theoretical radial velocity curves for our best-fit, long-term stable solutions with different period ratios. \textit{Top Panel}: Our overall maximum likelihood solution, which has $P_c/P_b \sim 7/5$. \textit{Middle Panel}: An example solution which shows clear libration of the 4:3 resonant angle. \textit{Bottom Panel}: Our maximum likelihood solution that also shows libration of the 3:2 resonant angle.}
\label{fig:rvt_comp}
\end{figure*}
\begin{figure}[htbp]
	\centering
	\includegraphics[ width=\linewidth]{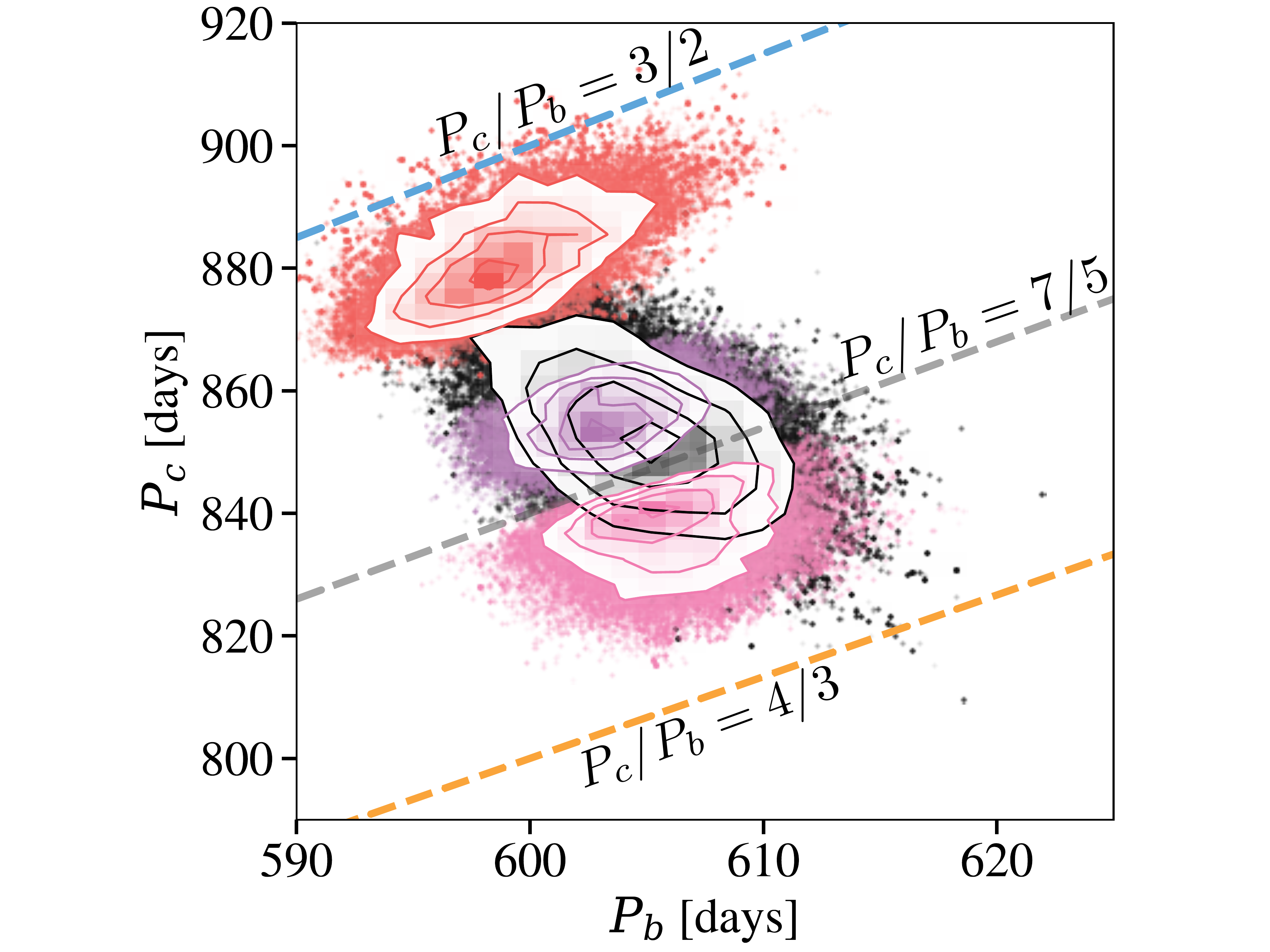}
		\caption{2D histograms of the posterior distributions for the planets' periods without long-term stability (black points, see Section \ref{fitting}) and the three modes identified for fits conditioned on stability for $10^6 \, P_c$ (pink, purple, and red points, see Section \ref{stab_fits}). Note that the plotted values refer to the periods at JD 2453213.895. Lines denoting exact ratios of $P_c/P_b$ are shown for ratios of 3:2 (blue), 7:5 (gray) and 4:3 (orange).}
\label{fig:per_ratio_comp_stab}
\end{figure}

\begin{figure}[htbp]
	\centering
	\epsscale{1.25}
	\plotone{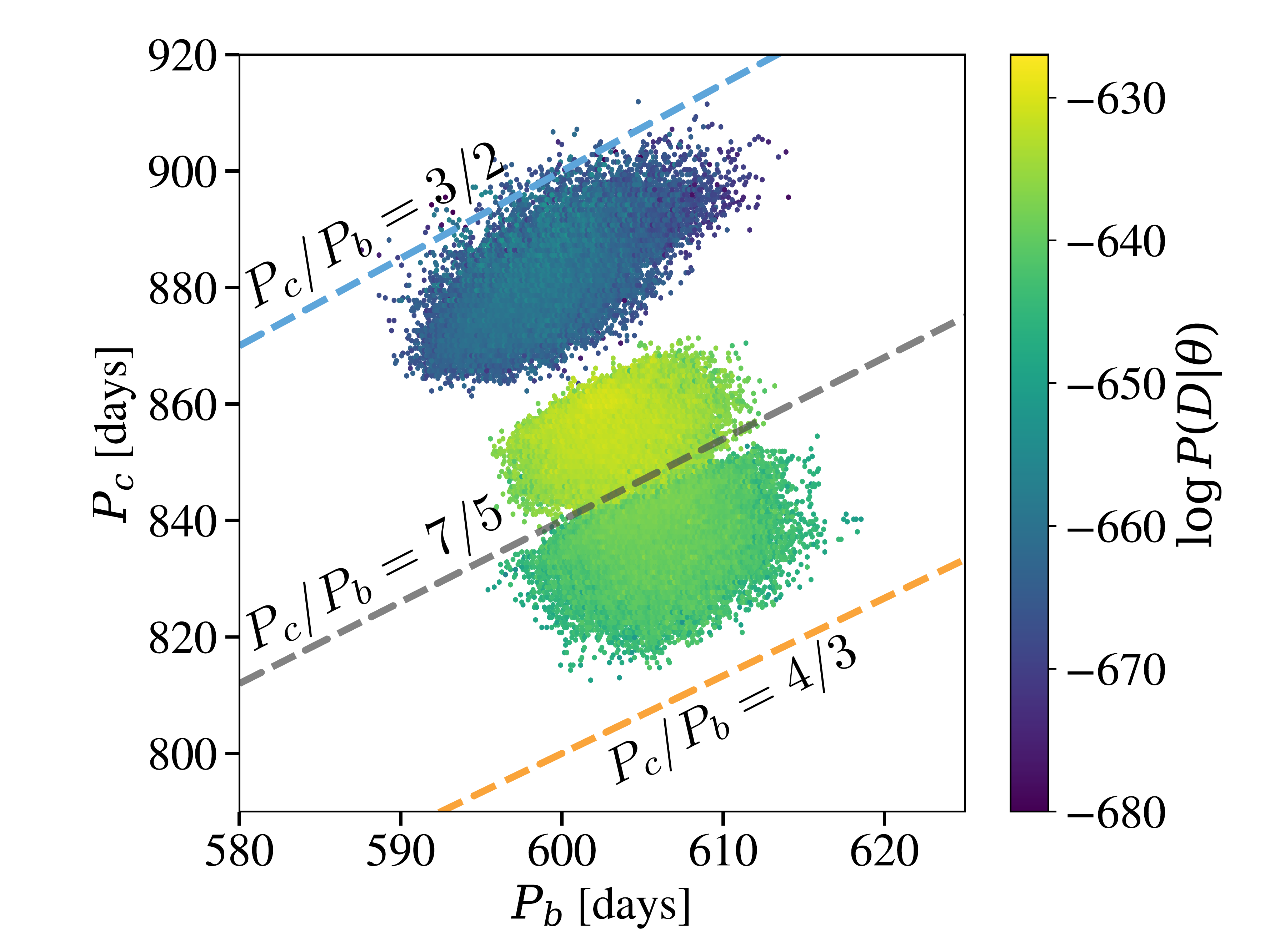}
		\caption{Posterior probability distributions shown in Figure \ref{fig:per_ratio_comp_stab}, with the points hexagonally binned and averaged. The points are colored by $\log P(D|\theta)$. The plotted values refer to the periods at JD 2453213.895. Note that the probability has not been properly marginalized over the other parameters, and is only meant to give a rough idea of the relative probability between the peaks (see text). Lines denoting exact ratios of $P_c/P_b$ are shown for ratios of 3:2 (blue), 7:5 (gray) and 4:3 (orange).}
\label{fig:blobs}
\end{figure}

As previously discussed, all three of these posteriors represent different modes of the overall posterior distribution of orbital parameters. A 2D histogram of the posterior distribution of $P_c$ vs. $P_b$ is shown in Figure \ref{fig:per_ratio_comp_stab}, overplotted with the non-stable 2D histogram. We use this plot to give an idea of where each modes lies in $P_c$ vs $P_b$ space; we stress that each mode is pulled from a separate posterior distribution, meaning that the relative likelihood of the modes is not indicated by the density of points in each 2D histogram. Given where each mode lies relative to the non-stable histogram however, it is clear that the mode at period ratios slightly larger than 7:5 will have the overall highest likelihood. To further emphasize this point, in Figure \ref{fig:blobs} we plot $P(D|\theta)$, i.e. the likelihood, hexagonally binned in $P_b$ vs. $P_c$ space and averaged. Again, we stress that this is not a proper marginalization over the other parameters in our space; however, since it can be seen in Appendix \ref{corner_plots} that the posterior distributions for the other parameters occupy similar regions of parameter space, this plot still gives a rough idea of the relative probability in each mode without being quantitatively rigorous.

Figure \ref{fig:blobs} makes it clear that the mode identified near 7:5 is by far the most likely -- it is higher in likelihood than the 4:3 by a factor of $\sim\exp(10-15)$, and the 3:2 mode by a factor of $\sim\exp(20-25)$. If we were concerned only with agreement between the data and our model, this mode would constitute our full posterior distribution. However, given the large shift in period ratio seen when more data is added to the RV signal, it is important to identify possible modes near the best-fit solution, as these modes may prove to be the ``true" solution when more data is added.

\begin{figure}[htbp]
	\centering
	\epsscale{1.25}
	\includegraphics[ width=\linewidth]{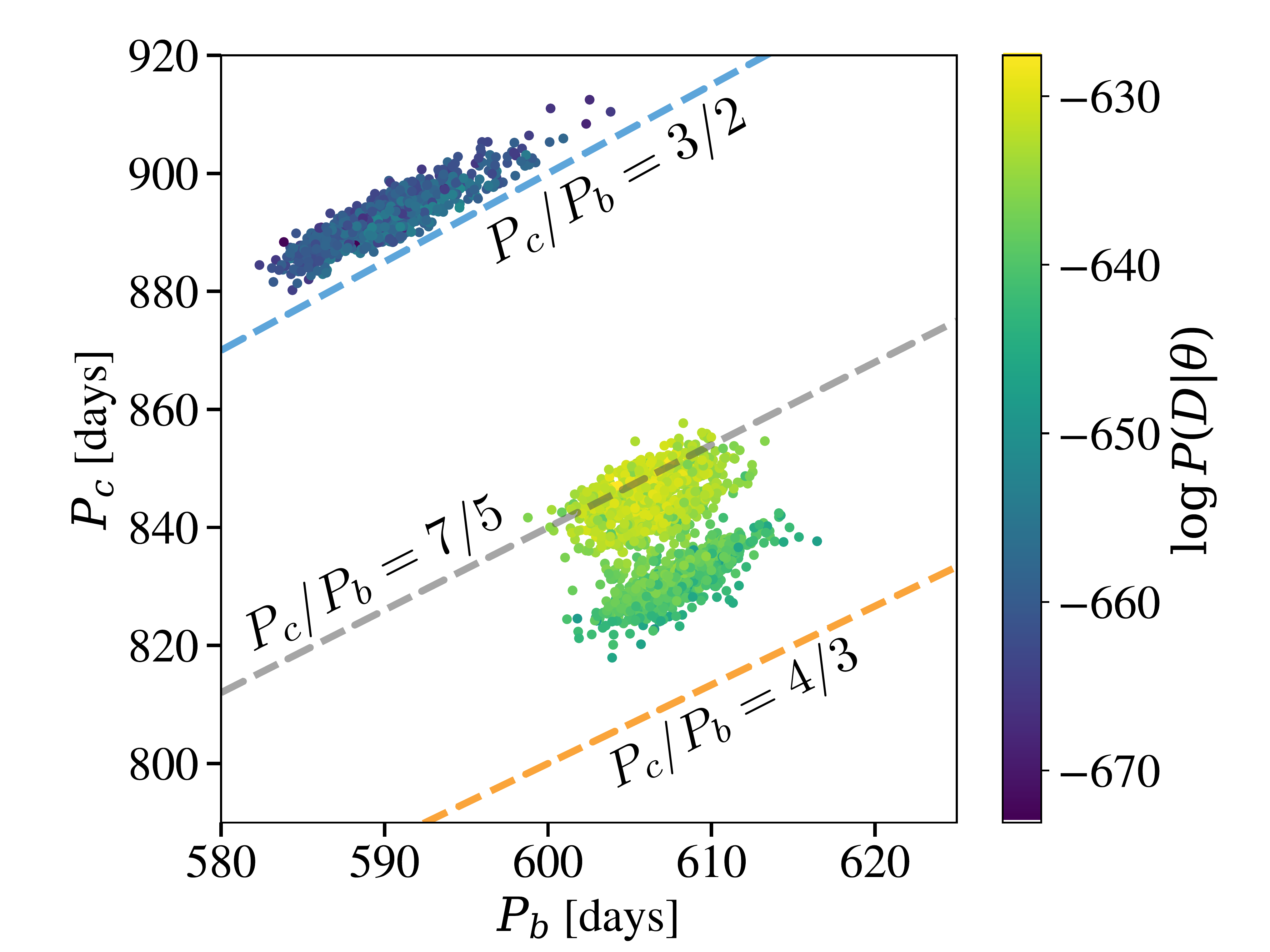}
		\caption{Osculating period values averaged over $500 P_c$ for 1000 draws from the three modes of the posterior distribution identified by our MCMC search. Each mode lies close to a different fixed value of $P_c/P_b$. Colors are the same as those in Figure \ref{fig:blobs}. Lines denoting exact ratios of $P_c/P_b$ are shown for ratios of 3:2 (blue), 7:5 (gray) and 4:3 (orange). See Section \ref{post} for a discussion.}
\label{fig:mean_per}
\end{figure}

Though we have identified three different modes of our posterior distribution clustered around three different values of the period ratio, all of the periods discussed thus far refer to the periods of the two planets at epoch; over time, the periods of the two planets will oscillate due to their mutual perturbations. To get a better sense of mean values of period ratio for these three modes, and to give a sense of the likelihood in each mode, we randomly sample 1000 points from each of our posterior distributions. For each point, we numerically integrate the system for $500 \, P_c$, and compute the mean values of $P_b$ and $P_c$. These values are plotted in Figure \ref{fig:mean_per}, along with $P(D|\theta)$ for each point. Integrating out the solutions has little effect on the period ratios for points in the 4:3 posterior---these points remain at values slightly larger than a 4:3 period ratio. For the 7:5 posterior however, the average periods all lie much closer to an exact 7:5 ratio, or slightly below, whereas their initial ratios were generally above 7:5. For the 3:2 points the ratios all now lie above 3:2, while their initial period ratios were all below. 

After this long term averaging over orbital elements, the period ratio distributions of our posterior modes lie even more clearly along or near lines of constant period ratio. This provides further support to the idea that these orbital configurations are stabilized by mean motion resonance. We explore this idea further in Section \ref{res}.

\subsection{Stable Regions of Period-Period Space} \label{dart}

\begin{figure}[htbp]
	\centering
	\epsscale{1.25}
	\includegraphics[trim=0 0 0 30, clip, width=1.1\linewidth]{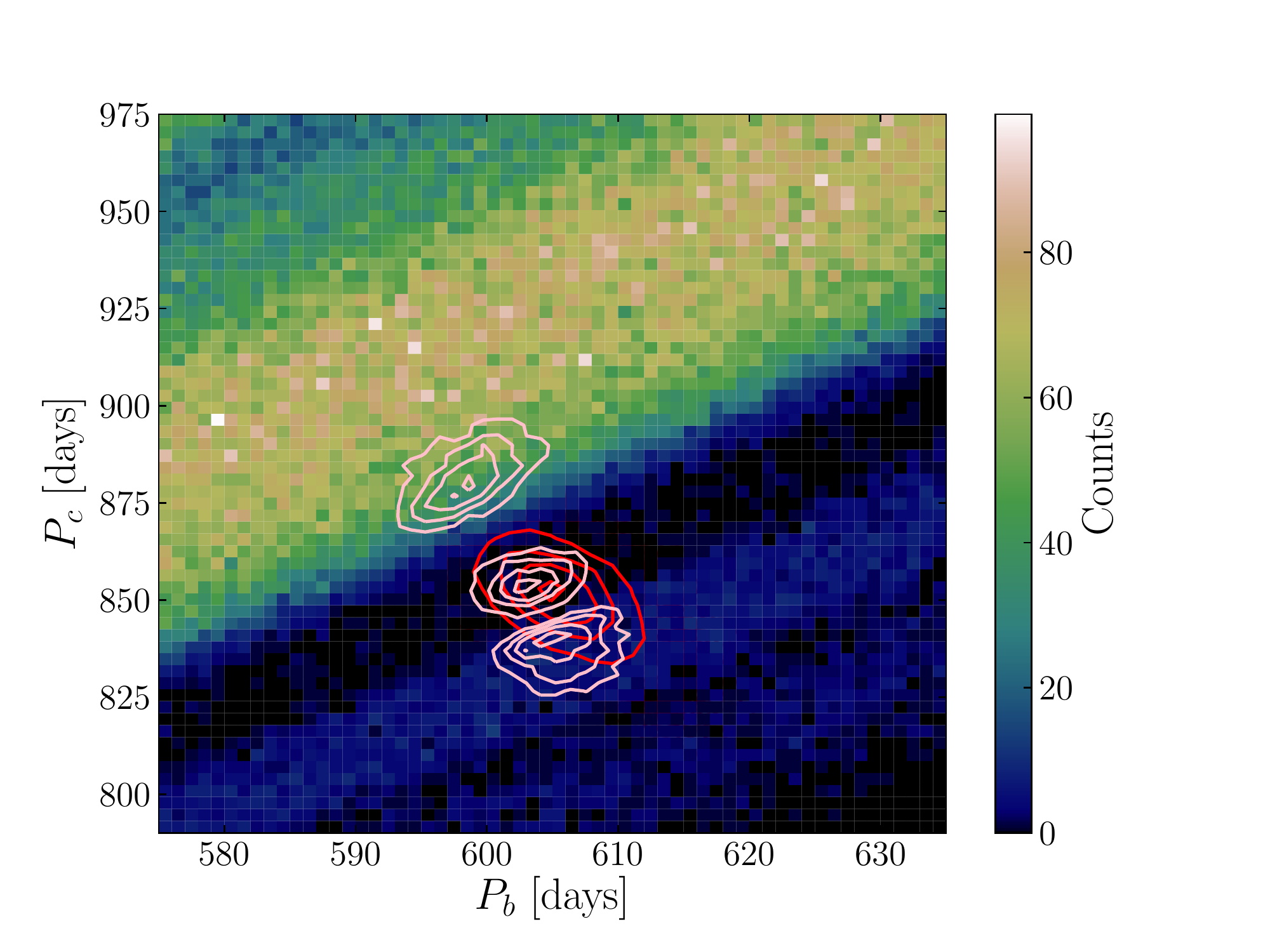}
		\caption{2D histogram of results from a Monte Carlo simulation of stable planetary systems. Orbital parameters for the two planets are randomly drawn, and systems that pass the stability criteria described in the text are recorded. The non-stable posterior distribution of the planetary periods is shown in red, and the three long-term stable modes are shown in pink.}
\label{fig:dart_throw}
\end{figure}

To check whether we have identified all of the possible modes, we preform a simple Monte Carlo simulation to analyze the stable regions near the planetary parameters we have identified. We initialize $10^6$ planetary systems, randomly drawing all parameters, except for the planetary periods, from normal distributions centered on the values for the parameters identified from the other modes. We used the following parameters for the normal distributions, where $\mu$ denotes the mean of the normal distribution and $\sigma$ the standard deviation: $\mu_{m_b} = 1.7 M_J$, $\sigma_{m_b} = 0.1$, $\mu_{m_c} = 1.2 M_J$, $\sigma_{m_c} = 0.2$, $\mu_{\lambda_b} = 300^\circ$, $\sigma_{\lambda_b} = 20$, $\mu_{\lambda_c} = 250^\circ$, $\sigma_{\lambda_c} = 40$, $\mu_{\log e_b} = -1.1$, $\sigma_{\log e_b} = 0.2$, $\mu_{\log e_c} = -1.5$, $\sigma_{\log e_c} = 0.1$, $\mu_{\omega_b} = 310^\circ$, $\sigma_{\lambda_b} = 100$, $\mu_{\omega_c} = 200^\circ$, $\sigma_{\omega_c} = 100$. The periods of the two planets are drawn from uniform distributions in the range 575 to 635 for planet b, and 790 to 925 days for planet c. Each planetary system is tested for stability in the manner described above, and the stable systems are recorded. A 2D histogram of the stable solutions in period space, along with the 3 modes and the non-stable posterior, are shown in Figure \ref{fig:dart_throw}.

Several features are apparent from Figure \ref{fig:dart_throw}. Firstly, stable regions of parameter space lie along diagonals running from the lower lefthand side to the upper right, indicating that stable regions of parameter space lie along regions of constant period ratio. Secondly, there is an extremely stable region of parameter space near the 3:2 MMR, and another stable region at ratios slightly larger than 4:3. Interestingly, the 7:5 mode, which has the overall highest likelihood, lies between these two stable regions. This is likely because the 7:5 mode is second order, making it weaker than the first order 3:2 and 4:3 resonances it is adjacent to. The lack of stable regions of parameter space near the 7:5 emphasizes the need to account for stability when considering the posterior probability distribution of the orbital parameters. From Figure \ref{fig:dart_throw}, it seems that if we are interested in additional possible stable modes, the only possibilities are the two regions near 3:2 and 4:3, which is precisely the other locations our search uncovered. Since any possible modes cannot lie too far from the non-stable posterior, Figure \ref{fig:dart_throw} provides further evidence that we have identified all relevant modes of the long-term stable posterior distribution.

\section{Analysis of Underlying Mean-Motion Resonance} \label{res}

As discussed in the last section, and demonstrated in Figure \ref{fig:mean_per}, the period ratios of the points in our posterior distribution lie near lines of constant period ratio, which supports the idea that these systems are in MMR. In order to investigate whether our stable best fit solutions are truly in resonance, we track the evolution of the resonant angle, $\phi$, over time. A $p/q$ MMR between a massive perturber and a massless test particle is characterized by libration of the angle

\begin{align} \label{eq:phi}
\phi = p \lambda_{\rm{outer}} - q \lambda_{\rm{inner}} - (p-q) \varpi_{\rm{test}}
\end{align}
(see e.g. \citealt{Murray1998}). For a truly massless test particle, if the semi-major axis ratio and initial angles are perfectly tuned, $\phi$ is constant; for values slightly off from this region, $\phi$ oscillates sinusoidally. In the case of HD 200964, libration of the resonant angle will be complicated by the large masses of both planets---not only are both planets of comparable mass, but in addition both planets are relatively massive compared to the central star. Thus, we do not expect libration of the resonant angle to be particularly ``clean.'' 

\begin{figure}[htbp]
	\centering
	\includegraphics[width=\linewidth]{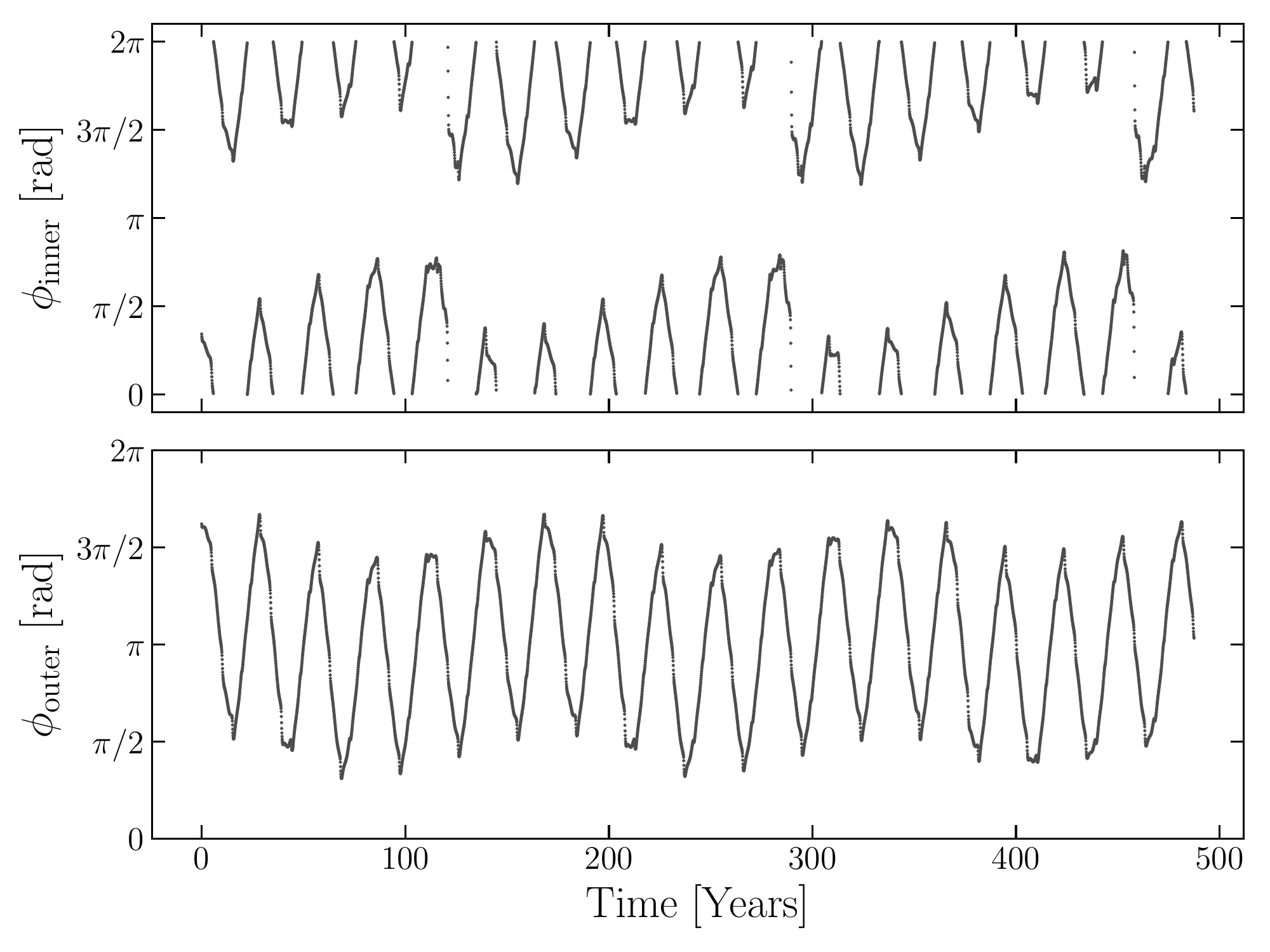}
        \epsscale{2.0}
		\caption{Value of the inner and outer 3:2 resonant angles for our best-fit 3:2 solution. Both angles clearly librate.}
\label{fig:phi_lib_3_2}
\end{figure}

We begin by discussing our solutions near a 3:2 period ratio, as they most clearly exhibit libration. The resonant angles for the maximum likelihood 3:2 solution are plotted in Figure \ref{fig:phi_lib_3_2}. The two resonant angles, $\phi_{\rm{inner}}$ and $\phi_{\rm{outer}}$, obtained by considering the inner and outer planets to be the test particle in Equation \eqref{eq:phi}, are shown. Both angles show clear libration, albeit with a large amplitude. Thus, it is straightforward to conclude that our long-term stable solutions near a 3:2 period ratio are in a 3:2 MMR.

\begin{figure}[htbp]
	\centering
	\includegraphics[width=\linewidth]{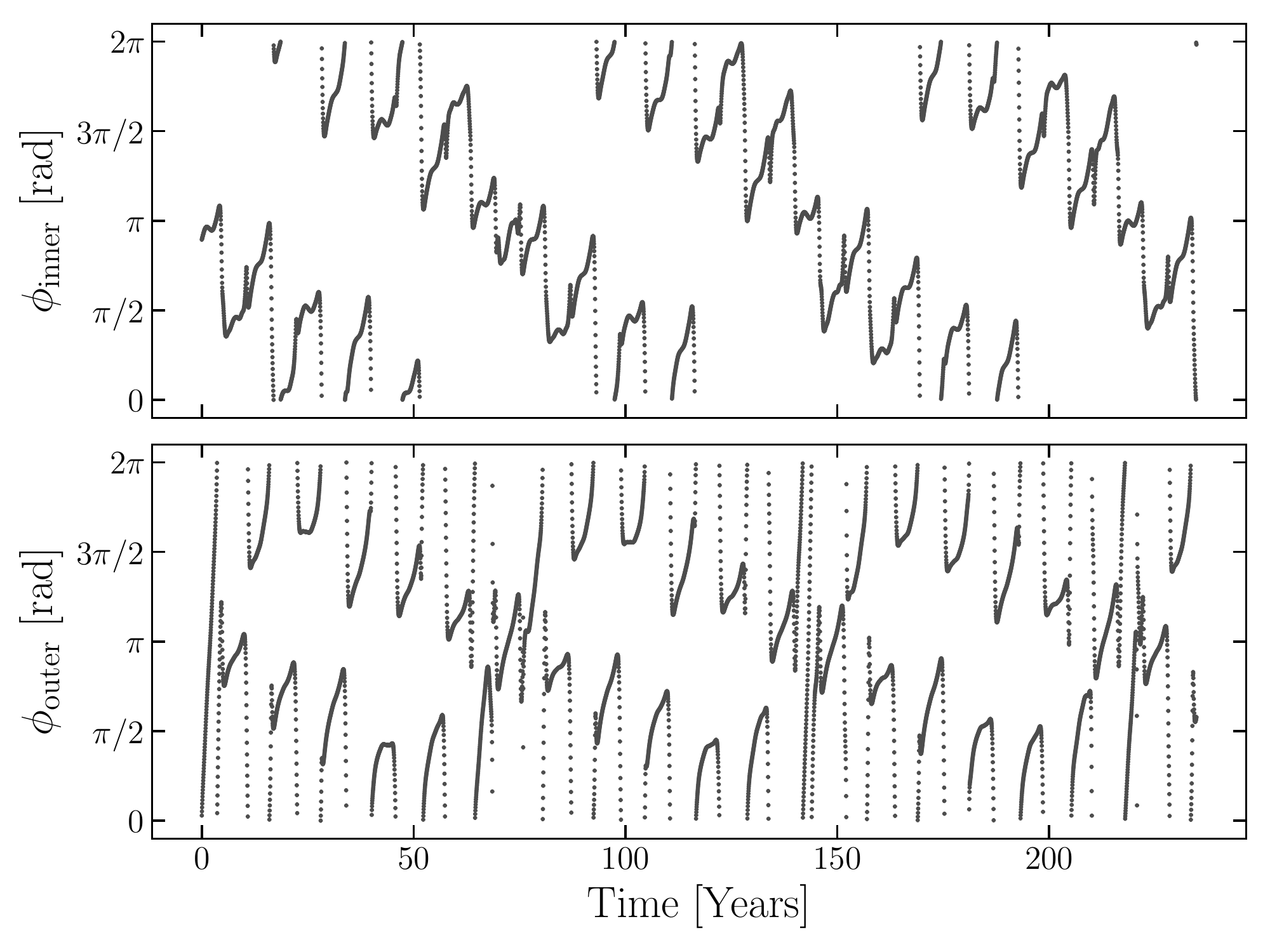}
        \epsscale{3.0}
		\caption{Value of the inner and outer 7:5 resonant angles for our best-fit solution, which are defined in Equation \eqref{eq:phi}. The angles do appear to show libration, but the large masses of both planets involved in the resonance complicate the libration pattern, as discussed in the text and demonstrated in Figure \ref{fig:phi_lib_ev}}.
\label{fig:bestfit_phi_lib}
\end{figure}

\begin{figure*}[htbp]
	\centering
\includegraphics[width=7in]{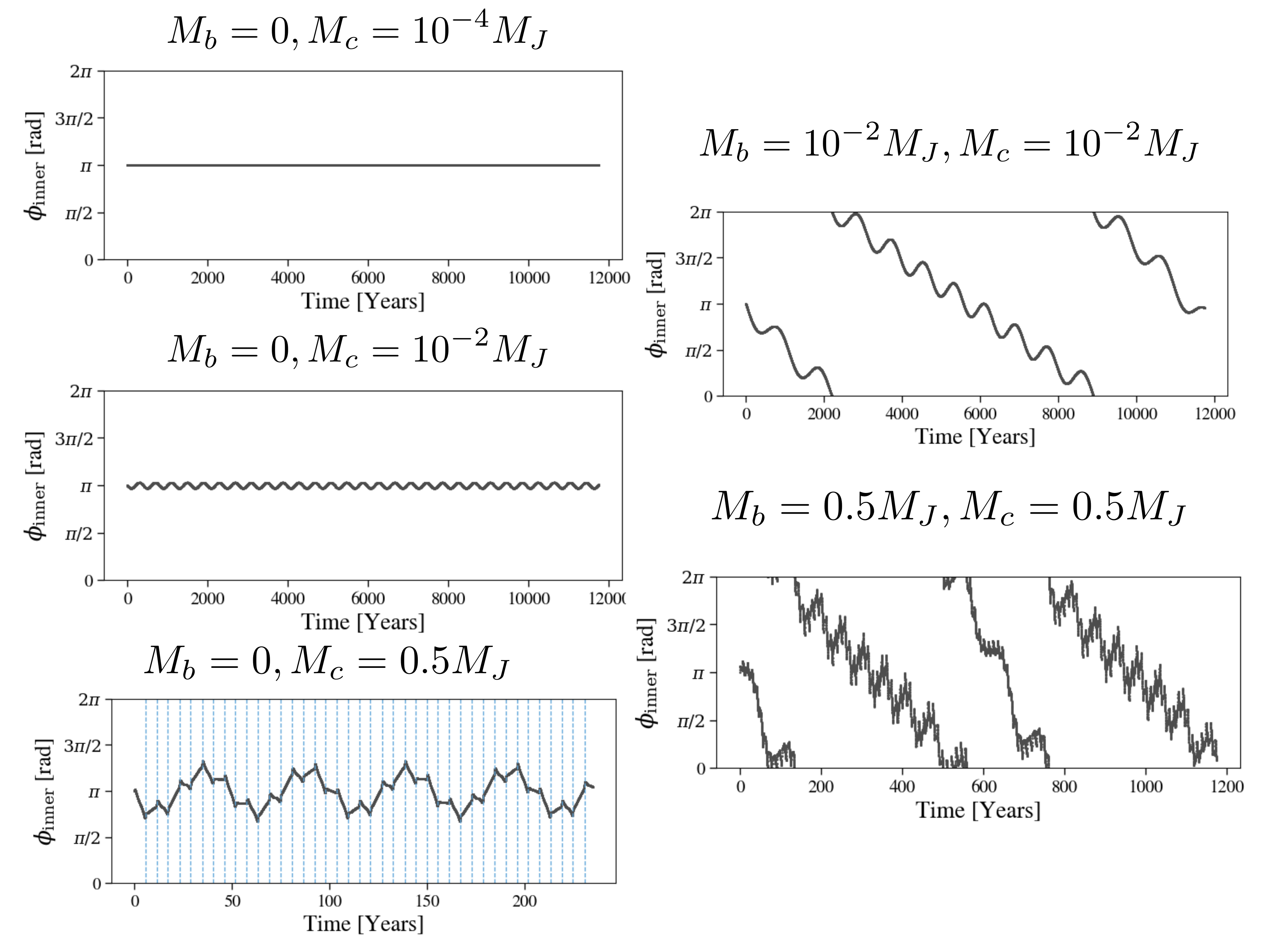}
		\caption{Evolution of $\phi_{\rm{inner}}$ for the 7:5 MMR as the masses of the planets involved in the resonance are increased. For low mass planets, libration of $\phi_{\rm{inner}}$ is easily discerned (middle left panel). As the mass of the perturbing planet is increased, kicks on a synodic timescale distort the libration pattern (bottom left panel; blue dashed lines denote conjunctions between the planets). If both planets have comparable mass, the center of the libration begins to circulate on the secular timescale (upper right panel). Finally, for large, comparably massive planets, both these effects serve to ``wash out" the libration of $\phi_{\rm{inner}}$ (lower right panel).}
\label{fig:phi_lib_ev}
\end{figure*}

For our 7:5 solutions however, the situation is more complex. The evolution of the 7:5 resonant angle for our maximum likelihood long term stable solution is shown in Figure \ref{fig:bestfit_phi_lib}. The two resonant angles, $\phi_{\rm{inner}}$ and $\phi_{\rm{outer}}$ are again shown. As can be seen in the figure, there does appear to be periodic variation in the value of $\phi$, but it is complicated by the presence of several other effects, which we enumerate in Figure \ref{fig:phi_lib_ev} by examining the evolution of $\phi$ as both the masses and the mass ratio of the planets involved in the resonance are increased. 

To begin, we plot the value of $\phi_{\rm{inner}}$ for two planets with $M_c = 10^{-4} M_J$, and $M_b = 0$. The angles of the planets are initialized such that the system begins perfectly in resonance. The planet's resonance angle is fixed at $\phi = \pi$ over the integration. As we increase the mass of the outer planet, the center of the resonance shifts off of an exact 7:5 period ratio. This causes the system to be initialized off resonance, causing $\phi$ to librate about $\pi$. Increasing the mass further to $0.5 M_J$ adds two new effects---firstly, the period of the libration of the resonant angle decreases dramatically, which is expected as the mass of the planets involved in the resonance increases. Secondly, there is now a much shorter period variation that has been introduced into $\phi$. This variation is caused by the outer planet perturbing the test particle during their closest approach, and therefore occurs on the synodic period of the planets. To illustrate this, we have noted conjunctions between the planets with dashed vertical lines. The strength of these synodic kicks makes the libration of the resonant angle less clear, though it can still be discerned by eye in this case. 

If we now give both planets comparable mass, as seen in the righthand top panel, the fact that the ``test" particle now has the same mass as the particle we are considering the ``perturber" for calculating $\phi$ causes the center of the libration to circulate as well, though the oscillation of $\phi$ about this circulating center can be clearly discerned.

Finally, we increase the mass of both planets to $0.5 M_J$. In this case, we see a combination of the two effects that were present previously---$\phi$ oscillates about a center that circulates, while the strong synodic kicks cause large oscillations of $\phi$ on a synodic period. 

These effects combine to produce the complicated behavior seen in the libration of $\phi_{\rm{inner}}$ for our best fit solution---for such high mass planets, the synodic kicks are extremely strong, and are on top of the rapid circulation of the center of the resonance. For contrast, in Appendix \ref{3_2_ev} we give analogous plots for the 3:2 resonant angle in Figure \ref{fig:phi_lib_ev_3_2}. In this case, the strength of the 3:2 resonance causes much less significant aberration from test particle case, even when both planets are $\sim M_J$. 

For the 4:3, we can find orbital configurations that show clear libration even for very massive planets. However, the orbital configurations that match the data well appear to be only marginally in the 4:3 resonance or not at all, since the complicated effects seen in $\phi$ are not due to the massive planets involved in the resonance alone. To illustrate this, in Appendix \ref{3_2_ev}, Figure \ref{fig:phi_lib_ev_4_3}, we plot the evolution of $\phi$ in a manner analogous to the plots made for the 3:2 and 7:5 MMRs. 

\begin{figure}[htbp]
	\centering
	\includegraphics[width=\linewidth]{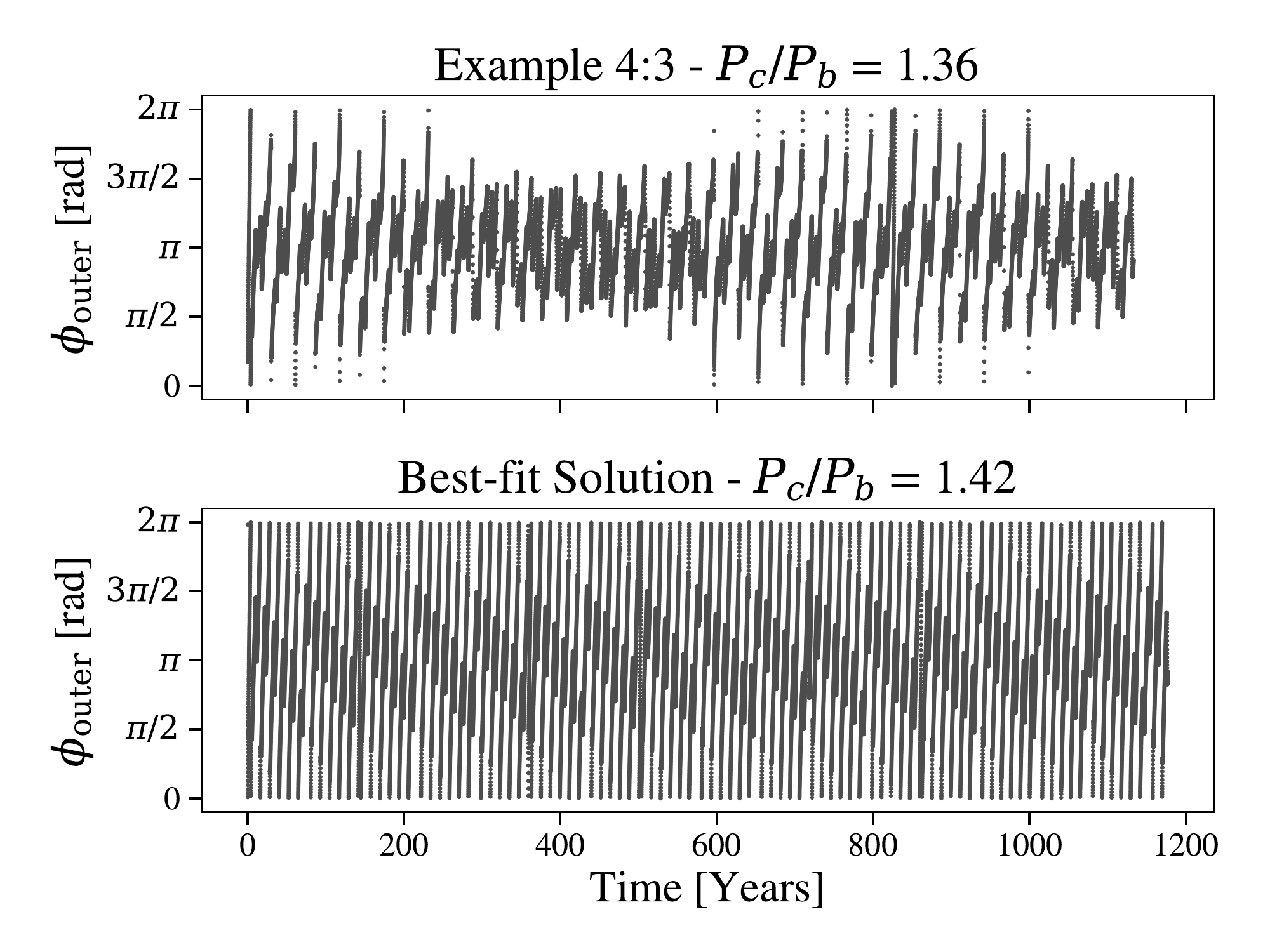}
        \epsscale{2.0}
		\caption{Value of the outer 4:3 resonant angle for two orbital configurations drawn from our posterior distribution. In the upper panel, we plot $\phi_{\rm{outer}}$ for a case where the period ratio of the planets is close to 4:3, and $\phi_{\rm{outer}}$ appears to librate. In the lower panel we plot the 4:3 outer resonant angle for our maximum solution; the angle appears to circulate in this case.}
\label{fig:bestfit_phi_lib_4_3}
\end{figure}

For the points in our posterior distribution, we only observed behavior similar to libration for the 4:3 resonance in $\phi_{\rm{outer}}$. An illustration of this is shown in Figure \ref{fig:bestfit_phi_lib_4_3}, which plots the outer resonant angle for a solution that does appear to show libration, and for our best-fit solution, which shows circulation. There appears to be a continuous evolution in behavior as the period of planet c is increased: for lower values of period, the outer 4:3 resonant angle does appear to librate about $\phi = \pi$, which is expected for a 4:3 MMR, though with a complex structure. For the larger period ratio solutions we find, i.e. those near a period ratio of 7:5, the angle appears to circulate instead. 

In summary, the 3:2 solutions we find are the only for which identification of the MMR through libration of the resonant angle is straightforward. For the 7:5 period ratio solutions, $\phi$ does appear to show periodic behavior which is clearly distinct from circulation. Interpretation of this behavior is not straightforward, though it does appear that the behavior of $\phi$ for the 7:5 MMR is consistent with libration for two Jupiter mass planets perturbing one another. For the 4:3 solutions, we see continuous behavior as the period ratio is increased, ranging from clear libration for period ratios closer to 4:3 to clear circulation for period ratios equal to or larger than 7:5.

\section{Reanalysis of Early Data} \label{reanaly}
Having now found several viable period ratios for long-term stable fits to the data, this now raises the question of whether the multiple resonances we have identified could have been found with just the data published in \citetalias{jphc11}. We therefore apply our methodology to just the Lick and Keck11 datasets, and analyze what aspects of the results we have presented can be found from those data sets alone.

To begin, we use a methodology similar to that discussed in Section \ref{stab_fits} to find a long term stable posterior distribution of orbital parameters. We perform initial optimization from several different locations in parameter space, including the parameters reported by \citetalias{jphc11}. We then run an initial $N$-body MCMC search from the best-fit obtained through optimization, without stability included, until we have a converged posterior distribution with $\sim 10^6$ points. At this point we perform a $10^6$ year rejection sample on our posterior, which leaves us with around 200 points in parameter space. This rejection sample identifies two clear regions of stability, one near a 4:3 period ratio and one near a 3:2. We follow up our rejection sampling with MCMC searches conditioned on stability starting in both of these regions. 

\begin{figure}[htbp]
	\centering
	\includegraphics[width=1.1\linewidth]{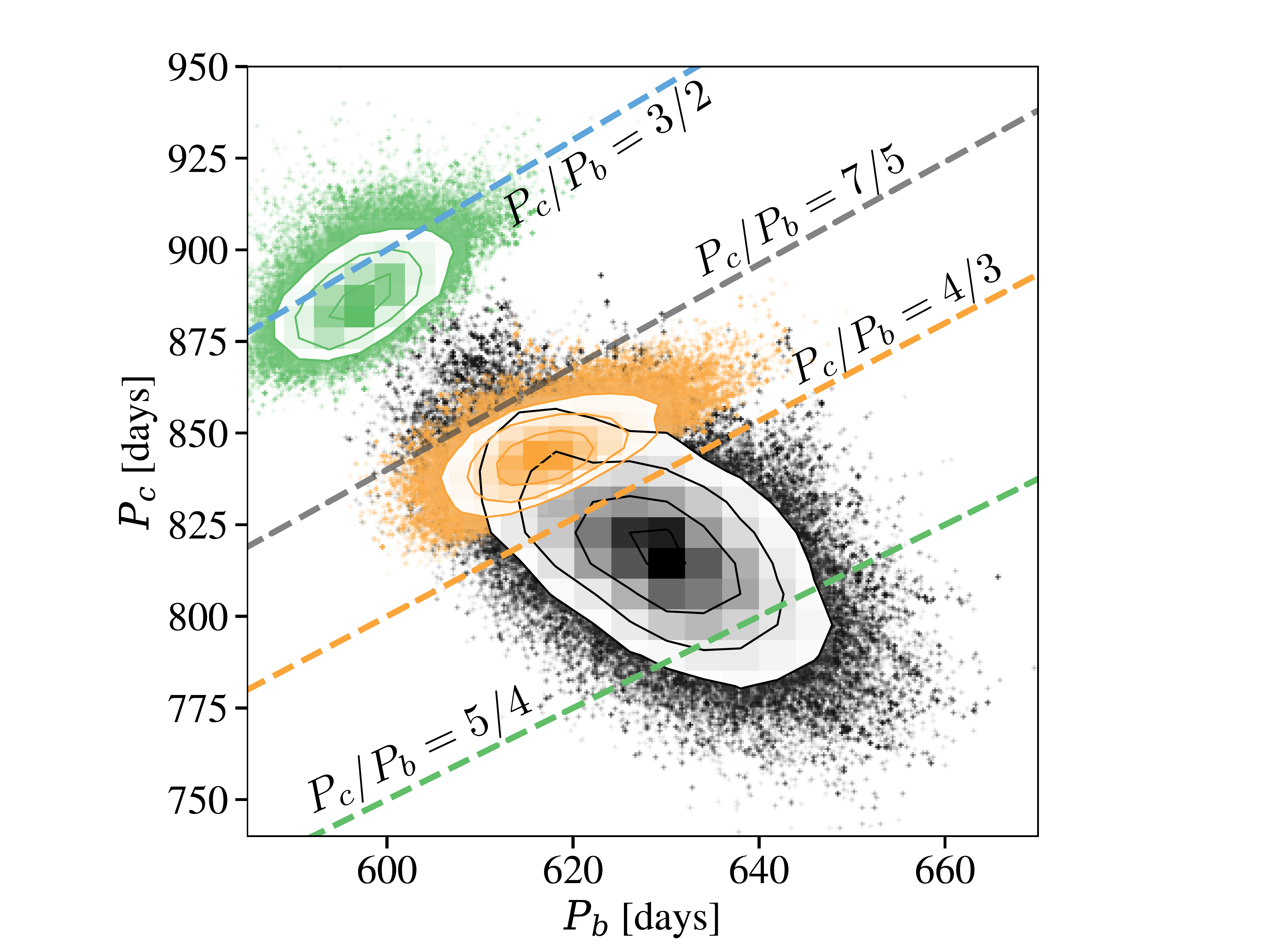}
		\caption{2D histograms of the posterior distributions for our planetary parameters from analysis of just the datasets used in \citetalias{jphc11}. The black points show the distribution without long-term stability, while the orange and green points show our posterior conditioned on stability for $10^6 P_c$. Lines denoting exact ratios of $P_c/P_b$ are shown for ratios of 3:2 (blue), 7:5 (gray), 4:3 (orange), and 5:4 (green).}
\label{fig:per_ratio_early}
\end{figure}

The resulting posterior distribution is shown in Figure \ref{fig:per_ratio_early}, along with the $N$-body integrated posterior without stability. As can be seen in the figure, the posterior distribution near 3:2 is quite similar to the one found for our longer dataset, while the 4:3 distribution is broader and at slightly larger values of $P_b$. It is notable here that the stable regions of parameter space are quite distant from the best-fitting region, which for the early data is at low values of period ratio. This result is in contrast to our analysis of the full data set, for which the best-fitting and stable regions lie on top of one another. This means that stability analysis is even more important when the data set is not as complete.

Thus, in addition to the 4:3 solution, we can identify the 3:2 orbital solution from analysis of the early data alone. However, it is interesting to note that the 7:5 solutions are not identified by this early search; it is only with the inclusion of more data that the 7:5 is even identified as a solution. 

\section{Possibility of a Third Planet} \label{third}
Though the two planet configurations we have identified provide plausible long-term stable fits to the data, it is still possible there are other planets in the system. We briefly investigate this possibility by adding a third planet to our model and investigate the resulting change in our maximum likelihood.

We initialize our fitting of third planet by looking at a periodogram of the residuals of our data. We take the strongest peak identified by the periodogram, which is is at $\sim$ 7 days, and use this orbital period as our starting point when adding the third planet. Given the low period, the residual could be due to a stellar signature. The rotation period of the star is likely to be too long to be causing this signature:    \citet{jps_2015} found a $v \sin i$ value of $1.88 \pm 0.23 \, \rm{km/s}$. Even at the upper end of the of the error bar, a simple calculation of rotation period using the value $R_{*}=4.92\,R_{\odot}$ gives $P_{{\rm rot}}=2\pi R_{*}/v\sin i\approx118\,{\rm days}$, which is clearly too long to give the $\sim$7 day planetary signal unless the star is rotating very close to pole on. On the other hand, the S-index values for HD 200964 do show some power at 8 days in the Keck data set, with a moderate correlation (Pearson correlation coefficient of 0.29), though this signal is not present in the APF data. Furthermore, there is significant power in both datasets around 26 days, which could likely be driving the correlation.  

Because the GLS favors a lower period for the third planet, it is unlikely that planet-planet interactions are important for modeling this third body. An initial optimization over the third planet's parameters further reinforces this point, as the optimization favors the third planet having low mass compared to the other two, with $M_d \sim 5 \times 10^{-2} M_J$. To enforce long-term stability in the system, we therefore fix the orbital parameters of planets b and c, and fit only the parameters of planet d. This means we will miss any covariances between the parameters of the hypothetical third planet and the two outer planets, but this method also ensures that the resulting three planet system exhibits long term stability.

We perform an MCMC search over the third planet's parameters, starting from the point identified by our optimization. The underlying parameter space is difficult to probe, with many of the solutions having log likelhioods that are comparable to the two-planet case. We do find orbital configurations that improve our log likelihood substantially enough that they may be significant. For a simple comparison we use a Bayseian information criterion (BIC) to compare our two models. We note, however, that the BIC is a surrogate for calculating the evidence, which is the more robust method (see e.g. \citealt{l_2007} for a discussion). For a given model, the BIC is calculated via
\begin{align}
    \text{BIC} = k \log n - 2 \log \hat{L}
\end{align}
where $\hat{L}$ is the maximum likelhiood, $n$ is the number of observations, and $k$ is the number of model parameters (i.e. 15 for the 2 planet case and 20 for the 3 planet case). In order to compare different models we calculate the BIC for each model and select the model with the lowest BIC.

\begin{figure}[htbp]
	\centering
	\plotone{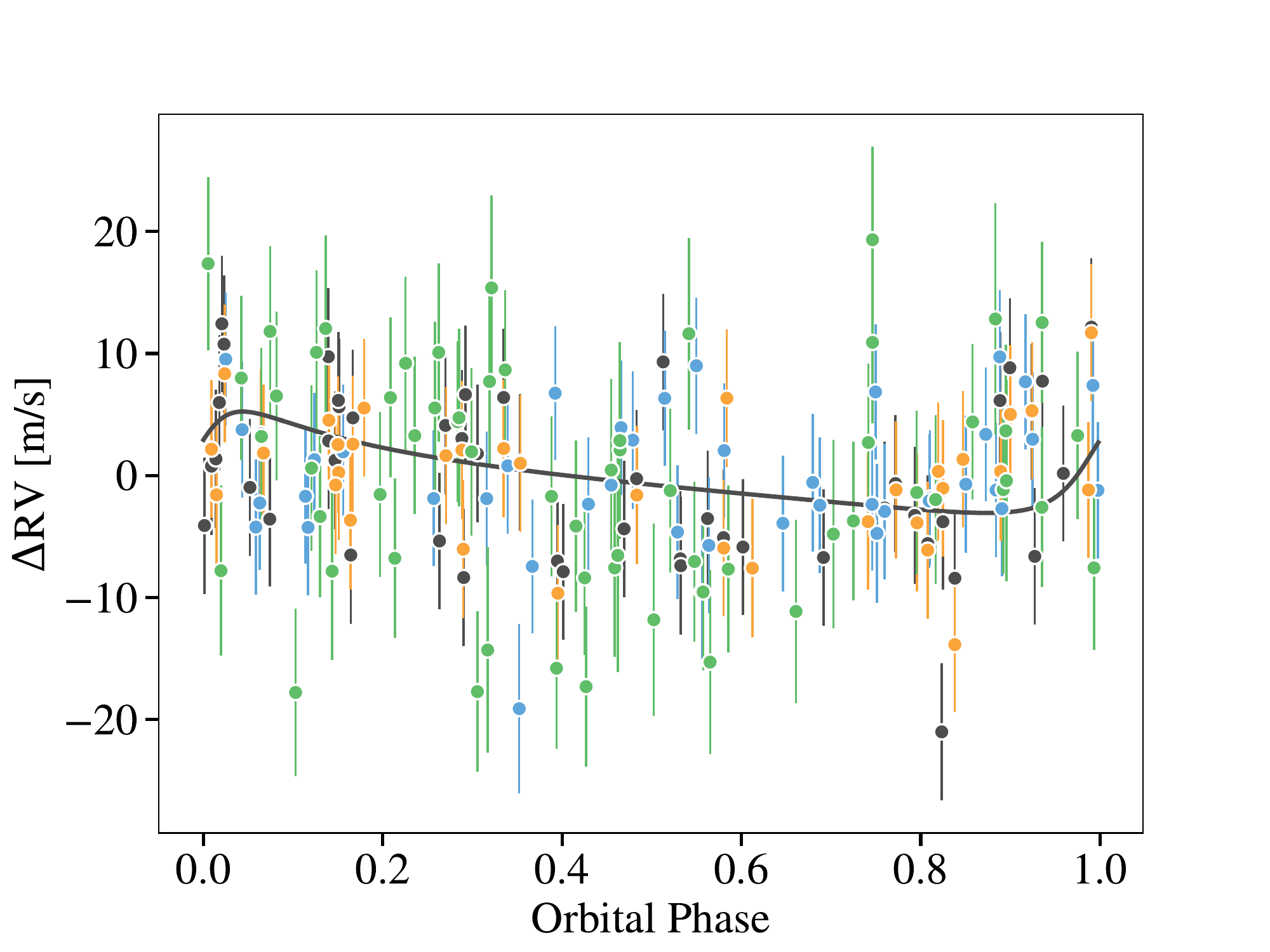}
        \epsscale{2.0}
		\caption{Radial velocity of the best-fitting third planet as a function of orbital phase. The radial velocity of planets b and c has been removed.}
\label{fig:planet_d}
\end{figure}

Our maximum likelihood third planet parameters are similar to those identified above: the planet is low mass ($M_d = 4.22 \times 10^{-2} M_J$), in a short period ($P_d$ = 7.89 days), highly eccentric ($e_d$ = 0.588) orbit. The $\Delta$BIC for this model versus our two planet model is $\Delta\text{BIC} = 4.90$. This means that the three planet model is preferred. The radial velocity signal for the third planet with the signals from planets b and c removed is plotted in Figure \ref{fig:planet_d}. Thus, while a three planet model does provide a smaller BIC, the BIC difference between the two models is not large, indicating that the three planet model is not strongly preferred over the two planet model. 

\section{Summary and Conclusions} \label{conc}
In this paper we have investigated the mean motion resonance between the two planets orbiting the star HD 200964. We find that the system is stabilized because it is in, or near, a mean motion resonance. However, which of three possible resonances the system is in (3:2, 4:3, or 7:5) remains unclear, as the full libration period of the system's resonance angle ($\sim$30 years) is longer than the observational baseline ($\sim$14 years). We also find indications of a possible ``low" mass ($M_p \sim 0.05 M_J$) third planet to the system on a short period ($\sim 8$ day) orbit, though this third planet is not strongly preferred over our two planet model.

Previous analyses (\citetalias{jphc11}) identified the system as being in a 4:3 resonance. By including stability in our searches we were able to identify additional long-term stable solutions near a 3:2 MMR, even using the same data analyzed in \citetalias{jphc11}, though 4:3 solutions remain better fits to this data set. Furthermore, by using radial velocity data spanning a longer timescale than previous works, we found that the best fitting orbital configurations were not in the 3:2 or 4:3 MMR, but instead had period ratios much closer to 7:5.  

The original identification of a 4:3 resonance was puzzling on theoretical grounds, as convergent migration of gas giants strongly prefers capture into the 3:2 rather than the 4:3 or 7:5. It is interesting to note that with the inclusion of more data the period ratio has gone up. We conclude that, this fact, along with the errors underlying the radial velocity measurements and the long timescale variation provided by libration of the resonant angle generate sufficient uncertainty in the period ratio that the 3:2 remains a plausible solution to the observed signal. 

If long period observations are not available, it is of paramount importance that long-term stability is included in fitting RV systems in MMR. For these shorter period solutions, the region of parameter space identified by simply finding the best-fit to the RV data can be a considerable distance from the regions of parameter space that exhibit long-term stability. Thus, requiring any proposed set of best-fit parameters to exhibit long-term stability is invaluable in identifying the the ``true" values of the underlying planetary system, which may be obscured by the strong perturbations of the planets on one another. 

\vspace{1mm}

\noindent The authors wish to thank Daniel Thorngren and Asher Wasserman for numerous insightful discussions on Bayesian statistics and interpretration of MCMC results. We would also like to thank the anonymous referee for their numerous helpful comments and suggestions, which greatly improved the quality of the manuscript. MMR and RMC acknowledge support from NSF CAREER grant number AST-1555385. WJG, BN, RMC, EC, NK, and JY also wish to thank the Heising-Simons foundation for supporting this work. JAB acknowledges support from MIT’s Kavli Institute as a Torres postdoctoral fellow. SSV gratefully acknowledges support from NSF grant AST-0307493. RPB gratefully acknowledges support from NASA OSS Grant NNX07AR40G, the NASA Keck PI program, and from the Carnegie Institution of Washington. 
The work herein is based on observations obtained at the W. M. Keck Observatory, which is operated jointly by the University of California and the California Institute of Technology, and we thank the UC-Keck and NASA-Keck Time Assignment Committees for their support. We also wish to extend our special thanks to those of Hawaiian ancestry on whose sacred mountain of Mauna Kea we are privileged to be guests. Without their generous hospitality, the Keck observations presented herein would not have been possible.  The work herein was also based on observations obtained at the Automated Planet Finder (APF) facility and its Levy Spectrometer at Lick Observatory. Computations were carried out using the Hummingbird computational cluster at UC Santa Cruz. Simulations in this paper made use of the REBOUND code which is freely available at \url{http://github.com/hannorein/rebound}.

\software{emcee \citep{fhl13}, REBOUND \citep{rl12}, Scipy \citep{scipy}, Pyevolve \citep{p_2009}, corner \citep{f16}.}

\appendix

\section{Posterior Distributions} \label{corner_plots}
\begin{figure*}
	\centering
	\plotone{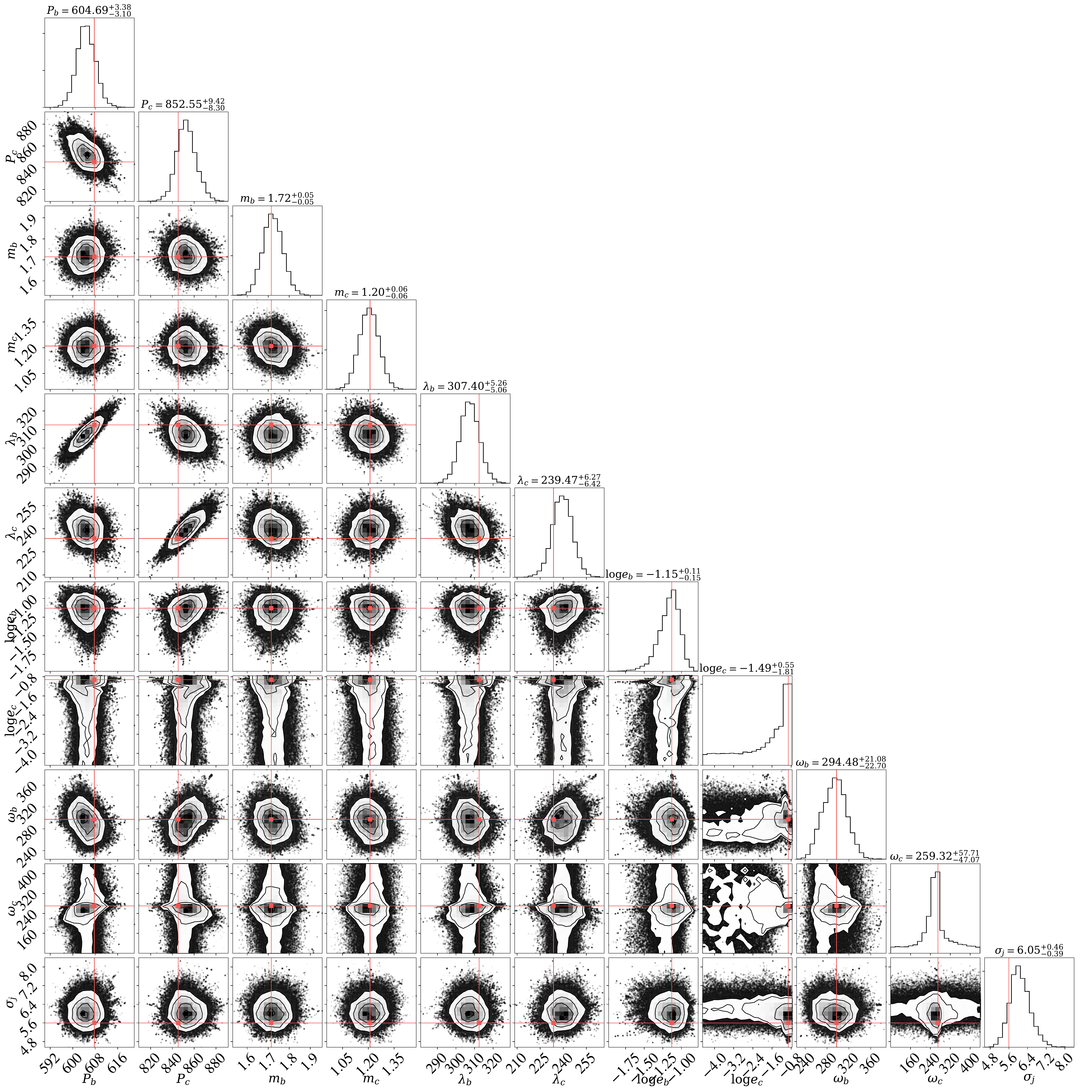}
		\caption{A corner plot showing the posterior distribution of the planetary parameters for the two planets orbiting HD 200964, without long term stability taken into account. All values for the orbital elements refer to the values at JD 2453213.895. The red lines indicate the location of the maximum likelihood parameters.}
\label{fig:nostab_corner}
\end{figure*}

\begin{figure}[htbp]
	\centering
	\plotone{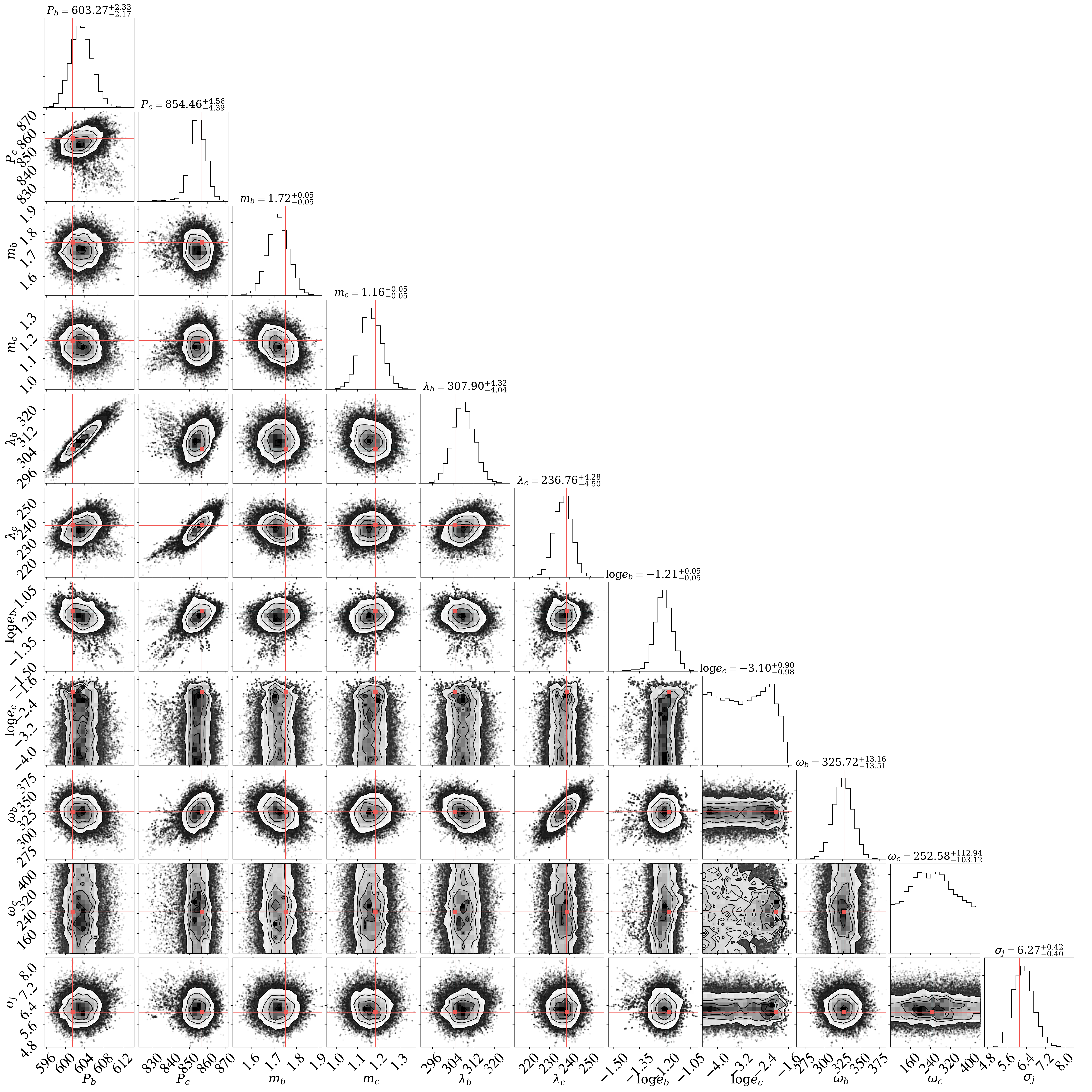}
        \epsscale{2.0}
		\caption{Corner plot showing the posterior distribution of parameters obtained by conditioning the likelihood function on stability for $10^6 P_c$. This posterior contains our best-fit, long-term stable solution. All values for the orbital elements refer to the values at JD 2453213.895. The red lines indicate the location of the maximum likelihood parameters.}
\label{fig:bestfit_stab_corner}
\end{figure}

\begin{figure}[htbp]
	\centering
	\plotone{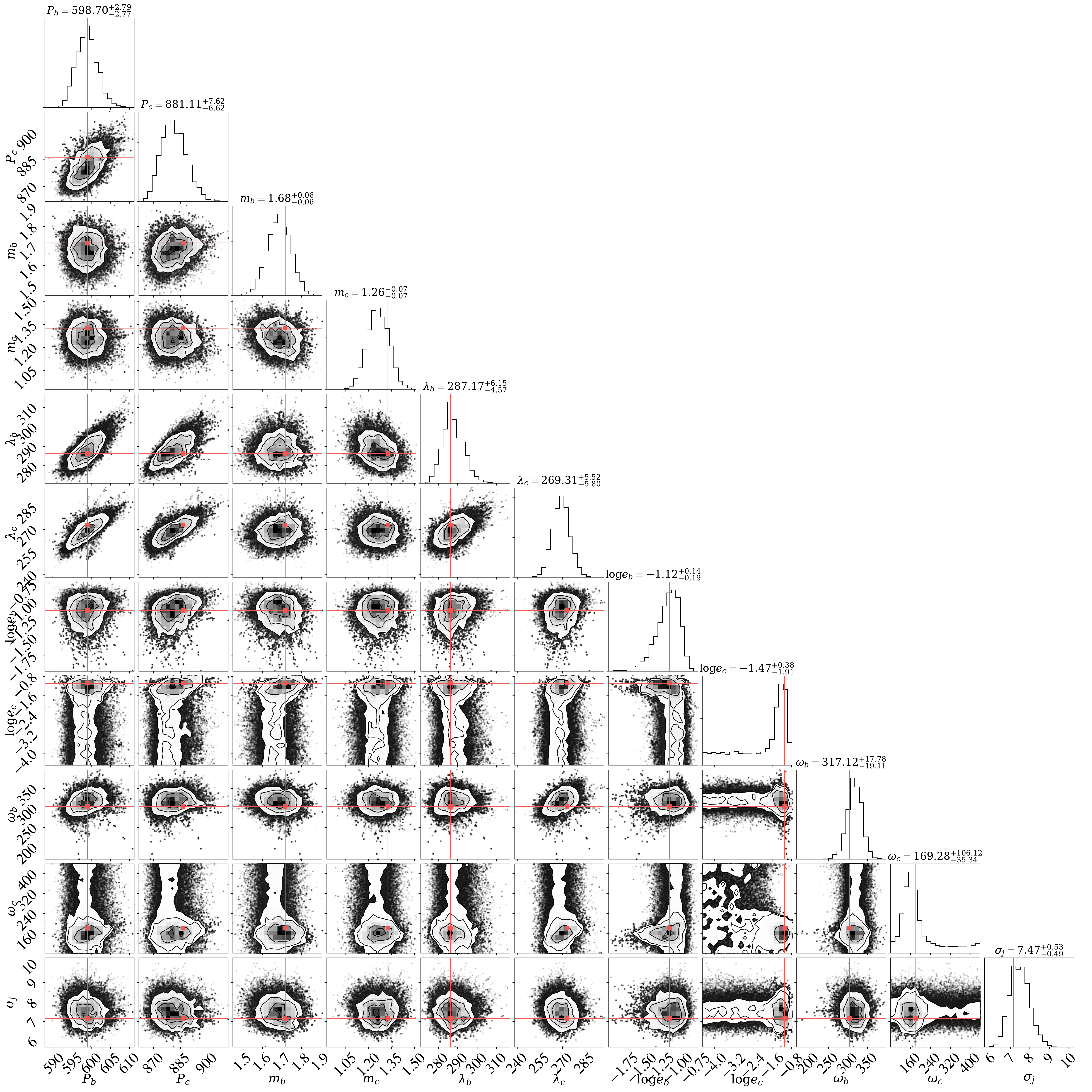}
        \epsscale{2.0}
		\caption{Corner plot showing the posterior distribution of parameters for period ratios $P_c/P_b \sim 3/2$, obtained by conditioning the likelihood function on stability for $10^6 P_c$. This posterior was obtained by initializing the search close to the 3:2 MMR. As can be seen in Figure \ref{fig:blobs}, the points in this posterior have an overall lower value of log likelihood than the posterior distribution shown in Figure \ref{fig:bestfit_stab_corner}.}
\label{fig:bestfit_stab_corner_3_2}
\end{figure}

\begin{figure}[htbp]
	\centering
	\plotone{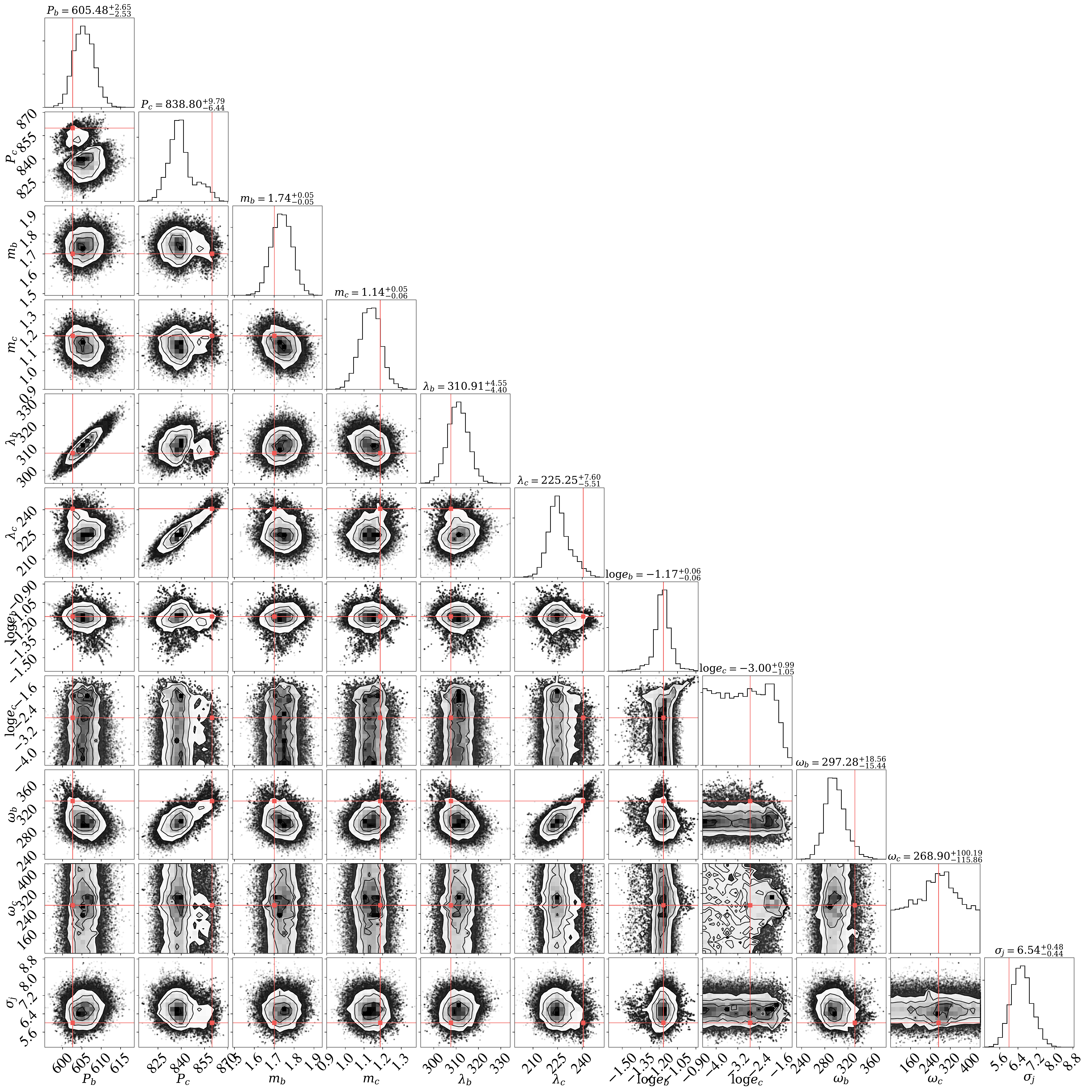}
        \epsscale{2.0}
		\caption{Corner plot showing the posterior distribution of parameters for period ratios $P_c/P_b \sim 4/3$, obtained by conditioning the likelihood function on stability for $10^6 P_c$. This posterior was obtained by initializing the search near previously published orbital solutions. This search identifies two clear modes in $P_c$ vs. $P_b$. Though the walkers spend more time at the lower period ratio mode, reinitializing the search at the higher period mode reveals that these solutions are a better fit to the data. See Section \ref{stab_fits} for more detail.}
\label{fig:bestfit_stab_corner_4_3}
\end{figure}

In this section we provide the full posterior distributions the solutions discussed in the text. Our best-fit posterior distribution without stability taken into account is given in Figure \ref{fig:nostab_corner}. Our best-fit posterior distribution conditioned on stability for $10^6 P_c$ is given in Figure \ref{fig:bestfit_stab_corner}, and our best-fit posteriors near a 3:2 and 4:3 period ratio are given in Figures \ref{fig:bestfit_stab_corner_3_2} and \ref{fig:bestfit_stab_corner_4_3}.

\section{Plots of the evolution of $\phi$ for the 3:2 and 4:3 MMR} \label{3_2_ev}

\begin{figure*}[htbp]
	\centering
\includegraphics[width=7in]{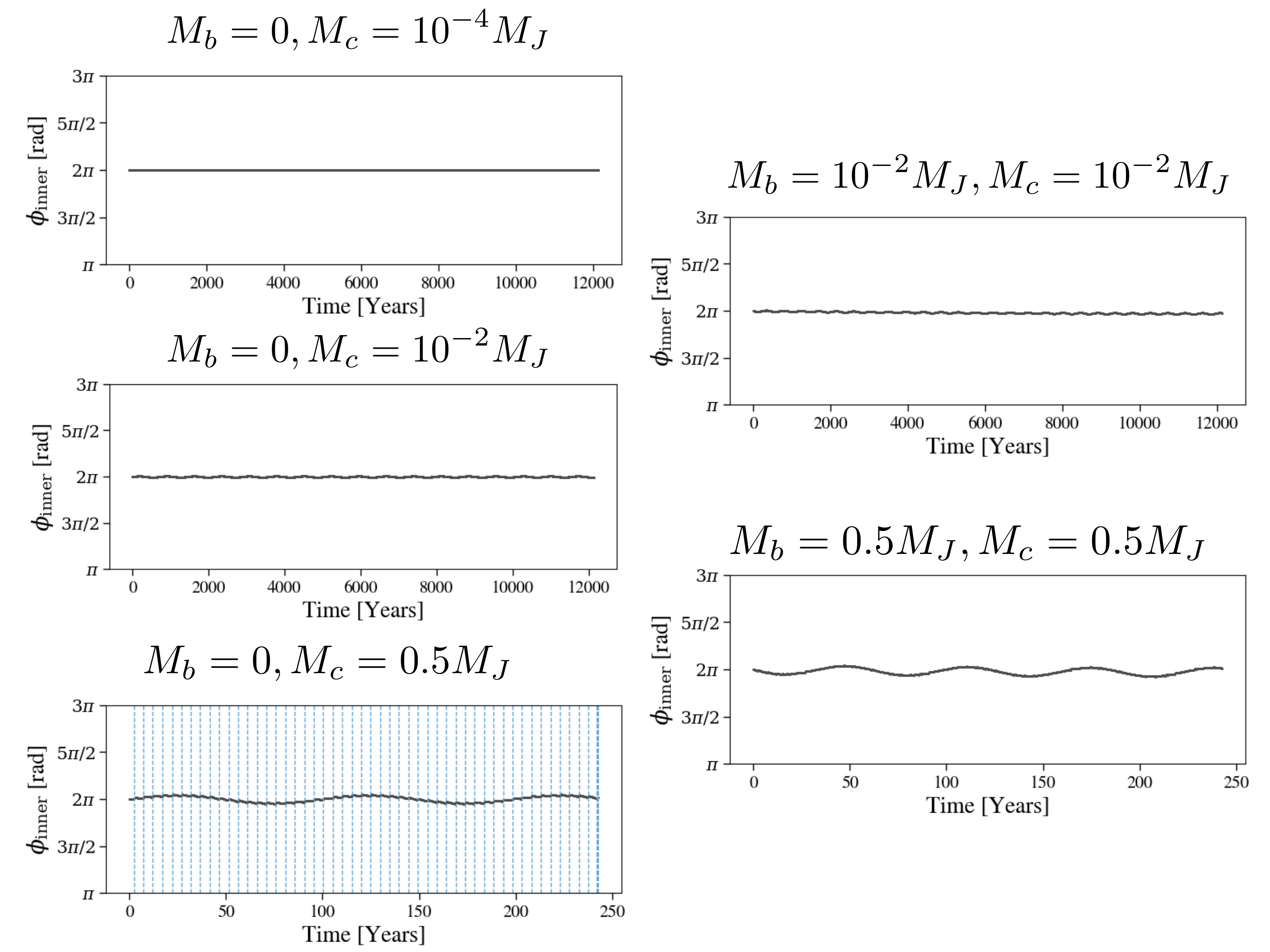}
		\caption{Evolution of $\phi_{\rm{inner}}$ for the 3:2 MMR as the masses of the planets involved in the resonance are increased. The panels are analogous to Figure \ref{fig:phi_lib_ev}}
\label{fig:phi_lib_ev_3_2}
\end{figure*}

\begin{figure*}[htbp]
	\centering
\includegraphics[width=7in]{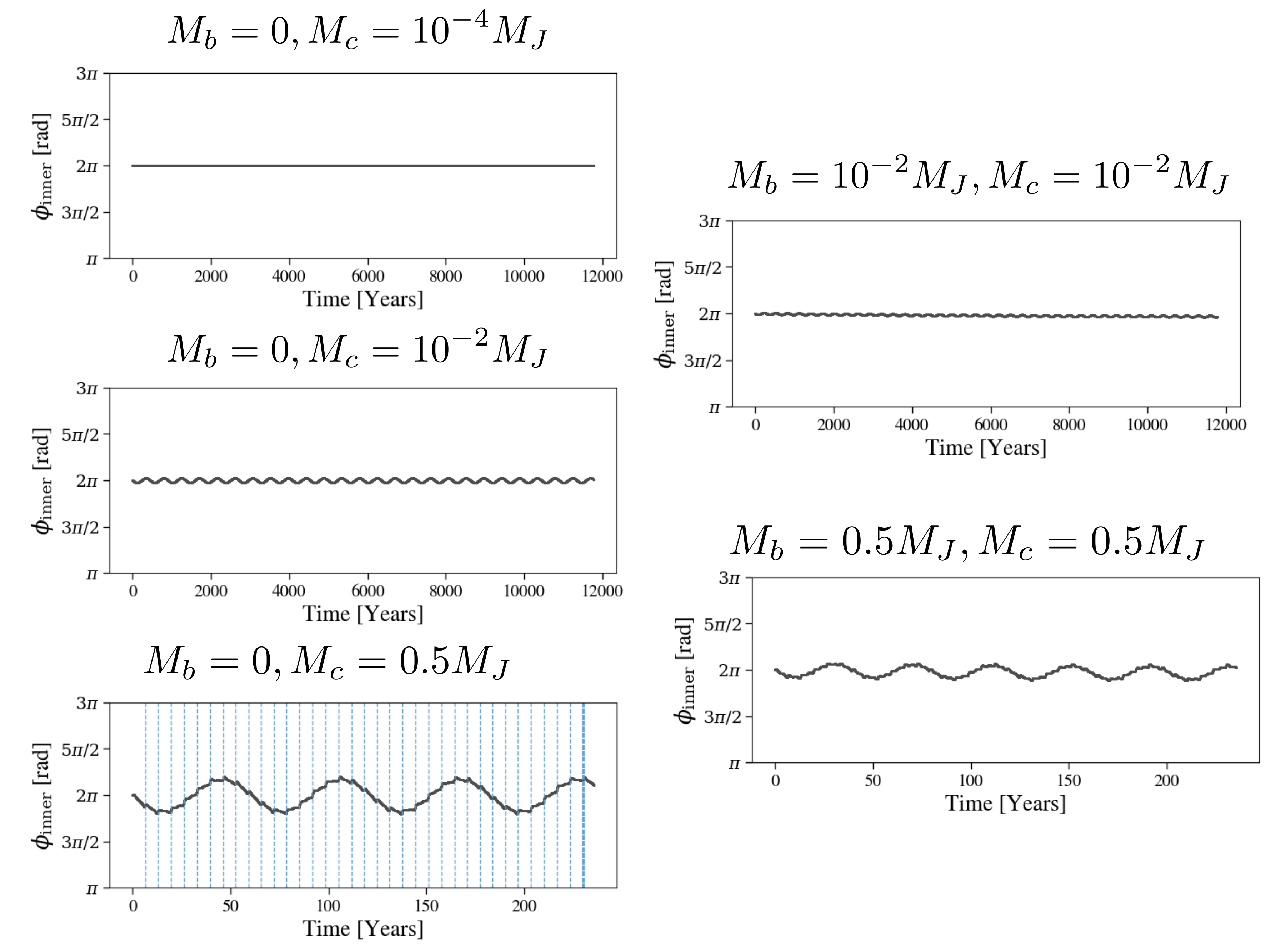}
		\caption{Evolution of $\phi_{\rm{inner}}$ for the 4:3 MMR as the masses of the planets involved in the resonance are increased. The panels are analogous to Figure \ref{fig:phi_lib_ev}}
\label{fig:phi_lib_ev_4_3}
\end{figure*}

In this section we make plots for the evolution of the resonant angle for the 3:2 and 4:3 MMR which are analogous to the ones plotted in Figure \ref{fig:phi_lib_ev}. The evolution of the 3:2 MMR is shown in Figure \ref{fig:phi_lib_ev_3_2}, and the evolution of the 4:3 MMR is shown in Figure \ref{fig:phi_lib_ev_4_3}.

\end{document}